\documentclass[pra,twocolumn,superscriptaddress,amsmath,amssymb,aps,floatfix]{revtex4-2}

\usepackage{graphicx}
\usepackage{dcolumn}
\usepackage{placeins}
\usepackage{bm}
\usepackage{amsmath,amssymb}
\usepackage{algorithm}
\usepackage{algpseudocode}
\usepackage{wrapfig}
\usepackage{setspace}
\usepackage[caption=false]{subfig}
\usepackage{ragged2e}
\DeclareCaptionJustification{justified}{\justifying}
\usepackage{amsfonts}
\usepackage{epsfig}
\usepackage{hyperref}
\usepackage{adjustbox}
\usepackage{rotating}
\usepackage{float}
\usepackage{dblfloatfix}
\usepackage{blindtext}
\usepackage{multirow}
\usepackage{array} 
\usepackage{bbold}
\usepackage{import}
\usepackage{physics}
\usepackage{verbatim}
\usepackage{listings}
\usepackage{xcolor}
\usepackage{xr-hyper}
\usepackage{cleveref}
\usepackage{newfloat}
\usepackage{caption}
\usepackage{subcaption}
\DeclareFloatingEnvironment[name={Fig S},fileext=lof]{suppfigure}
\usepackage[margin=0.6in]{geometry}
\begin{document}
\title{Phase vs coin vs position disorder as a probe for the resilience and revival of single particle entanglement in cyclic quantum walks
}
\author{Dinesh Kumar Panda}
\email{dineshkumar.quantum@gmail.com}
\author{Colin Benjamin}
\email{colin.nano@gmail.com}
\affiliation{School of Physical Sciences, National Institute of Science Education and Research Bhubaneswar, Jatni 752050, India}
\affiliation{Homi Bhabha National Institute, Training School Complex, Anushaktinagar, Mumbai
400094, India}
\begin{abstract}
Quantum states exhibiting single-particle entanglement (SPE) can encode and process quantum information more robustly than their multi-particle analogs. Understanding the vulnerability and resilience of SPE to disorder is therefore crucial. This letter investigates phase, coin, and position disorder via discrete-time quantum walks on odd and even cyclic graphs to study their effect on SPE. The reduction in SPE is insignificant for low levels of phase or coin disorder, showing the resilience of SPE to minor perturbations. However, SPE is seen to be more vulnerable to position disorder. We analytically prove that maximally entangled single-particle states (MESPS) at time step $t=1$ are impervious to phase disorder regardless of the choice of the initial state. Further, MESPS at timestep $t=1$ is also wholly immune to coin disorder for phase-symmetric initial states. Position disorder breaks odd-even parity and distorts the physical time cone of the quantum walker, unlike phase or coin disorder. SPE saturates towards a fixed value for position disorder, irrespective of the disorder strength at large timestep $t$. Furthermore, SPE can be enhanced with moderate to significant phase or coin disorder strengths at specific time steps. Interestingly, disorder can revive single-particle entanglement from absolute zero in some instances, too. These results are crucial in understanding single-particle entanglement evolution and dynamics in a lab setting.
\end{abstract}
\maketitle

\twocolumngrid
\textcolor{brown}{Introduction.--} Single-particle entanglement or SPE refers to entanglement between different degrees of freedom (DoF) of a single particle such as a laser-cooled ion trap or a photon~\cite{wineland,aqs,p2,aqs,fang,gratsea_lewenstein_dauphin_2020}. Quantum states exhibiting SPE can encode more extensive quantum information (QI) than their multi-particle analogs exploiting large Hilbert space~\cite{qjoining}. These SPE states are more robust to decoherence and advantageous for quantum communication and cryptography~\cite{aqs,p2,p4}. Further, these have applications in large-scale QI processing, quantum state exploration, and studying the interaction between fundamental particles ~\cite{aqs,gmeuse,gmeuse20,gmeuse2,qudit1,qudit2}. However, these are challenging to generate and control. Such states with large or maximal SPE can be generated efficiently via discrete-time quantum walks (DTQWs)~\cite{p2,fang,p4,p1,p3,qw93}. Note that DTQWs have experimental realization using lattice-based quantum systems whose position space matches the discrete lattice sites, such as cold atoms~\cite{cold1,karski},
photons~\cite{bian,ph1,ph2,ph3,ph4}, trapped ion~\cite{ion1,ion2} and NMR system~\cite{nmrqw}.

In a recent letter, we found recurrent generation of maximally entangled single-particle states (MESPS or maximal SPE states) via a single coin in cyclic DTQW (without disorder), see Ref.~\cite{p2}. However, how such MESPS (i.e., a state with maximal SPE) or an SPE state would behave in the presence of disorder is yet to be studied. In addition, whether a disorder can revive SPE is a fundamental research question. Such a study is paramount as disorder/noise is ubiquitous and sometimes unavoidable in quantum systems. We bridge this fundamental research gap and comprehensively investigate the effects of phase, coin, and position disorder (of arbitrary strengths) on DTQW dynamics in odd or even cyclic graphs and the generated position-coin (quNit-qubit) SPE states.

In a cyclic DTQW (CQW) without disorder (i.e., clean-CQW), one can generate MESPS for arbitrary initial states under certain conditions. We showed this in Ref.~\cite{p2}, and with a Hadamard coin operator, one generates recurrent MESPS with period four at time steps $t=1,5,9,...$, via the clean-CQW for an arbitrary initial state, see~\cite{p2}. In this letter, we find that the reduction in SPE is insignificant for low levels of phase or coin disorder, showing the resilience of SPE to minor disturbances. SPE is, however, more vulnerable to position disorder. For phase disorder, MESPS (maximal SPE state) at time step $t=1$ remains unperturbed for any arbitrary initial state. For coin disorder, however, only two specific phase-symmetric initial states give foolproof immunity against disorder for MESPS at $t=1$. Position disorder breaks \textit{odd-even} parity~\cite{sasha} (i.e., position $x=0$ probability: $P(x=0)\ne0$ at odd $t$) and distorts the physical \textit{time cone} of the quantum walker, unlike phase or coin disorder. SPE saturates towards a fixed value for position disorder, irrespective of the disorder strength, for more significant time steps.

Further, SPE can be enhanced with moderate or more significant phase or coin disorder strength at specific time steps. Notably, disorder can also revive single-particle entanglement. One can tune the SPE value by controlling the strength of various disorders~\cite{prl}, and this work aligns with lab implementations~\cite{bian,ph1,ph2,ph3,ph4} for quantum systems evolving via CQWs. This counterintuitive result of enhancement, revival and resilience of SPE is a byproduct of the interplay between disorder, interference and unitary quantum evolution~\cite{noise1,noise2,noise3,noise4}. This motivates strategies to harness noise/disorder for enhancing entanglement, safeguarding quantum coherence, and exploring quantum error mitigation (which is reliant on precise noise characterization) techniques for large-scale QI processing and quantum computation~\cite{noise1,noise5,noise6,noise7}.


Below, we present a generalized CQW model (with and without the disorder) for odd- and even-site cyclic graphs that beget a rich class of quantum states and generate SPE between the walker's position-coin DoF. Then, we discuss the behavior of generated SPE (from clean CQW dynamics) under phase disorder, coin disorder, and position disorder of arbitrary strengths. After that, we analyze the disorder effects on the generated SPE states, compare our results with other relevant approaches, and finally draw conclusions from this comprehensive study. Additional details on generalized disordered CQW dynamics, its properties, and SPE vs disorder for both 4- and 3-cycles are provided in Supplementary Material (SM)~\cite{supple}.

\begin{figure}[h!]
\includegraphics[width=7.4cm,height=1.85cm]{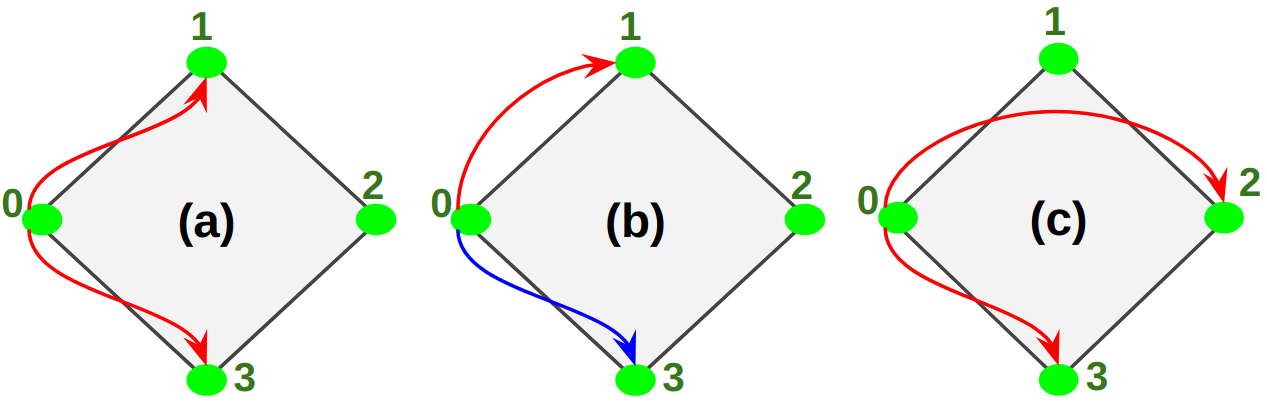}
\caption{(a) Clean-CQW; (b) CQW with phase or coin disorder, where amplitudes in coin basis change; (c) CQW with position disorder where jump length in shift changes randomly. Sites are shown as green dots, and red (for the same amplitudes) and blue (for different amplitudes) curves denote amplitudes of the particle on 4-cycle.}
\label{f1}
\end{figure}

\textcolor{brown}{\textit{CQW and single particle entanglement.--}}
A quantum particle evolving via DTQW dynamics on a $k$-site cyclic graph ($k$-cycle, see Fig.~\ref{f1}) is defined on a Hilbert space ($\mathcal{H}$), combining its $k$-dimensional position ($\mathcal{H}_P$) and 2-dimensional coin ($\mathcal{H}_C$) spaces. $\mathcal{H}_C$ and $\mathcal{H}_P$ are spanned by computational bases $\{\ket{0_c},\ket{1_c}\}$ and  $\{\ket{x_p}: x \in {0,1,2,\dots,k-1}\}$ respectively, and $\mathcal{H} = \mathcal{H}_P \otimes \mathcal{H}_C $. A physical quantum walker is a photon (or a cold-trapped ion) wherein $\mathcal{H}_C$ and $\mathcal{H}_P$ correspond to its polarization (hyperfine electronic states) and path (local position) DoF respectively~\cite{wineland,karski,bian,viera}.
When the quantum walker (particle) is initially localized at site $\ket{0_p}$ in an arbitrary superposition of coin states $\ket{q_c}$, its initial state,  $\ket{\psi(t=0)}$ reads,
\begin{equation}
\ket{\psi(t=0)}=\ket{0_p}\otimes[\cos(\frac{\theta}{2})\ket{0_c} + e^{i\phi}\sin(\frac{\theta}{2})\ket{1_c}],
\label{eq1}
\end{equation}
where $\theta \in [0, \pi]$ and $\phi \in [0, 2\pi)$. The quantum walker evolves unitarily via a time-dependent evolution operator $U_{k}(t)$ and after $t$ time steps, is in the evolved state,
\begin{align}
\begin{split}
\ket{\psi(t)}=& U_{k}(t)U_{k}(t-1)...U_{k}(1) \ket{\psi(t=0)},\\
=&\sum_{x=0}^{k-1}[\alpha_{0}(x,t)\ket{x_p,0_c} + \alpha_{1}(x,t)\ket{x_p,1_c}],
\end{split}
\label{eq2}
\end{align}
where the amplitudes $\alpha_{0}(x,t)\text{ and } \alpha_{1}(x,t)$ depend on initial state (Eq.~(\ref{eq1})) and the design of the evolution operator, $U_{k}(t) = \hat{S}(t)\cdot[\sum_{x}\ket{x_p}\bra{x_p}\otimes \hat{C}(\rho(x,t))]\;$. 
\noindent
We use a general qubit-coin operator,
\begin{equation}
\hat{C}(\rho) =
\begin{pmatrix}
\sqrt{\rho} & \sqrt{1-\rho}\\
\sqrt{1-\rho} & -\sqrt{\rho}
\end{pmatrix},
\text{with $0\leq\rho\leq1$,}
\label{eq3}
\end{equation}
\noindent to introduce arbitrary superposition in the walker's coin subsystem. Subsequently, a shift operator $\hat{S}(t)$ is employed to move the walker to the left (right) in the position-basis by hopping (jump) length $J(t)$ which is contingent on the coin state $\ket{0_c}$ ($\ket{1_c}$). The typical translation (or shift) operator reads,
\begin{align}
\begin{split} \hat{S}(t) =& \sum_{s=0}^{1}\sum_{x=0}^{k-1}\ket{((x+2s-J(t)) \text{ mod } k)_p}\bra{x_p}\otimes\ket{s_c}\bra{s_c}.
\end{split}
\label{eq4}
\end{align}
To obtain the well-known Hadamard walk in a $k$-cycle, one sets $\rho=\frac{1}{2}$ (i.e., Hadamard coin $ \hat{H}=\hat{C}(\frac{1}{2})$ ) and $ J(t)=1$, $\forall t$, see also Ref.~\cite{p2}. We observe for the Hadamard coin CQW and any arbitrary general initial state Eq.~(\ref{eq1}), which is  clean-CQW (i.e., without disorder), the following three fundamental properties:

1).~It preserves the \textit{odd-even} parity~\cite{sasha} only for even $k$-cycle graphs like 4-cycle, i.e., at even (odd) $t$, only even (odd)-numbered sites have finite probabilities of walker's occupation. For details, see SM Sec.~A. Physically, this implies a photon (a walker) can not be found at even sites once it has evolved to an odd $t$.

2).~It respects the \textit{physical time-cone} of the walker for all $k-cycle$, i.e., a walker found at position $\ket{x_p}=\ket{m_p}$ for time step $t$, will have finite probability of being found at positions up to $\ket{x'_p}=\ket{((m\pm n)\text{ mod } k)_p}$  at time-step ($t+n$). For details, see SM Sec.~A.

\begin{figure}[h!]
\includegraphics[width=9.8cm,height=4.5cm]{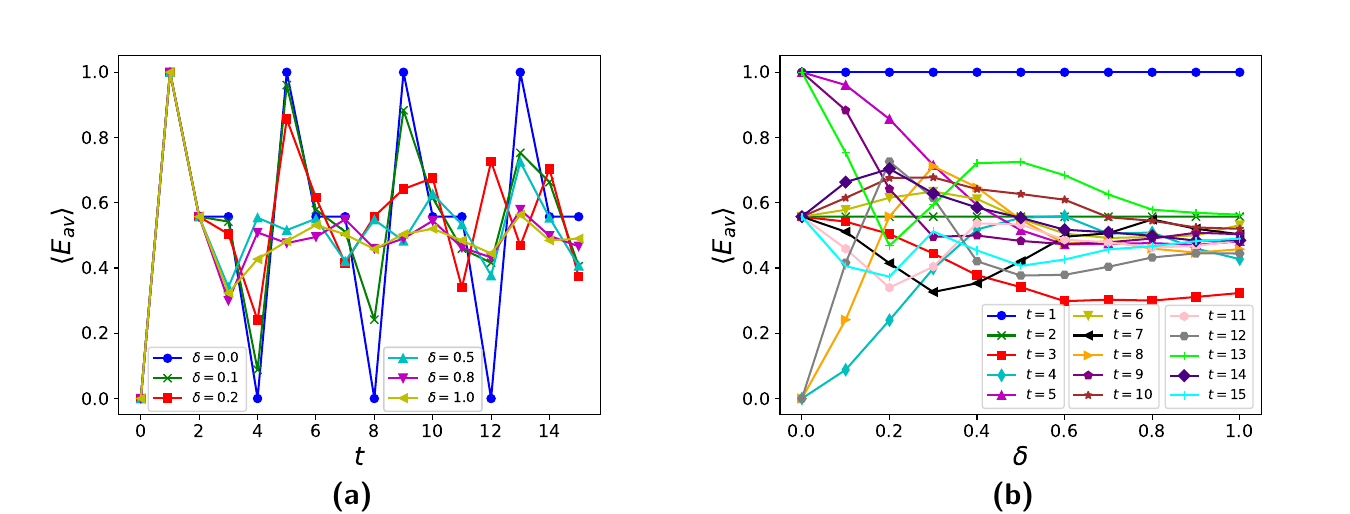}
\caption{(a) SPE $\langle E_{av} \rangle$ vs. time step $t$ for different static-phase-disorder strengths ($\delta$); (b) SPE $\langle E_{av} \rangle$ vs. static-phase-disorder strength $\delta$ for different $t$ for 4-cycle.}
\label{f3}
\end{figure}

3).~It generates MESPS at time step $t=1$ for any arbitrary $k$-cycle graph. Corollary: for 4 and 8-cycles, MESPS is generated recurrently with period 4 at time step $t=1,5,9,...$, see Fig.~\ref{f3}(a) (blue), see also~\cite{p2}. The generic time-evolution of CQW dynamics with general initial state Eq.~(\ref{eq1}) for arbitrary qubit-coin (Eq.~(\ref{eq3})) and general translation operator (Eq.~(\ref{eq4})), which generates high-dimensional (quNit-qubit) single particle entanglement between the position (quNit, $N=k$) and the coin (qubit) DoF of the walker, is detailed in SM Sec.~A.

In order to quantify the amount of quNit-qubit entanglement (SPE) in the disordered CQW, we employ entanglement entropy ($E$)~\cite{p2,janzing_2009,viera}. $E$ at any $t$ is defined as the von-Neumann entropy of the reduced density matrix $\rho_c(t)=\text{Tr}_p(\ket{\psi(t)}\bra{\psi(t)})$, i.e., $E(\rho_c)=$ $-$Tr($\rho_c$log$_{2}$$\rho_c$). If  $E_{0}$ and $E_{1}$ are the eigenvalues of $\rho_c(t)$, one equivalently estimates, $E = -\sum_{s=0}^{1}E_{s}\text{log$_{2}$}E_{s} \;$.
The average SPE is then,
$E_{av} = \frac{1}{\pi} \int_{0}^{\pi} E\;d\theta\; $ for $\phi=\frac{\pi}{2}$ and SPE states are guaranteed for $0<E_{av}\le1$. For separable states like the initial state (\ref{eq1}), $E_{av}=0$, and for the state evolving via a clean-CQW at $t=1$(see SM Sec.~A and Eq.~(2) with Hadamard coin), we see $E_{av}=1$, i.e., it is a MESPS.

\textcolor{brown}{\textit{Phase disorder and SPE in CQW.--}}
Phase disorder in the CQW setup can be site-dependent (static) or time-dependent (dynamic). Below, we first investigate the effect of static and then the dynamic phase disorder on CQW evolution and generated SPE.

\textcolor{brown}{\textit{Static-phase disorder.--}}
The static-phase disorder is introduced in the CQW setup through the following translation operator (modified Eq.~(\ref{eq4}) without time-dependence), i.e.,\newline
\begin{align}
\begin{split} \hat{S}_{sp} = \sum_{s=0}^{1}\sum_{x=0}^{k-1}\ket{((x+2s-1) \text{ mod } k)_p}\bra{x}\;e^{i\alpha_s(x)}\otimes\ket{s_c}\bra{s_c},
\end{split}
\label{eq8}
\end{align}
where $\alpha_s(x)$ with $s\in\{0,1\}$ are site-dependent phase angles. Any CQW property will only depend on the fluctuations in relative phase difference $\alpha_0(x)-\alpha_1(x)$, as we can factor out the phase $e^{i\alpha_1(x)}$ (while estimating a physical observable), i.e.,\newline
$\hat{S}_{sp}= \sum_{x=0}^{k-1}e^{i\alpha_1(x)}(\ket{((x-1) \text{ mod } k)_p}\bra{x}\;e^{i(\alpha_0(x)-\alpha_1(x))}\\\otimes\ket{0_c}\bra{0_c}+\ket{((x+1) \text{ mod } k)_p}\bra{x}\;\otimes\ket{1_c}\bra{1_c}).$\newline
Thus, randomizing the phase angle $\alpha_0(x)$ only, i.e., by choosing it randomly from the set $[0,\delta]$ in units of $\pi$ with a uniform distribution~\cite{ntuyong}, and keeping $\alpha_1(x)=0$ (constant), introduces a finite random phase difference, i.e., a generalized phase disorder in the CQW setup. The probability density function for a uniform distribution reads,
\begin{equation}
f(\alpha_0(x)) = \frac{1}{b - a} \quad \text{for real} \quad a \leq \alpha_0(x) \leq b .
\label{eq9}
\end{equation}
Here, $a=0, b=\delta$ and the full CQW evolution is then, $U_{k}(t) = \hat{S}_{sp}\cdot[\sum_{x}\ket{x}\bra{x}\otimes \hat{H}]\;$. Clearly, at $\delta=0$, we retrieve the clean-CQW. Thus,
$\ket{\psi(t)}=\hat{S}_{sp}\cdot[\sum_{x}\ket{x}\bra{x}\otimes \hat{H}]\ket{\psi(t-1)}\;.$

In the numerical simulation of the static-phase disorder case, a set of random on-site phase $\alpha_0(x)$ values is generated, with its size corresponding to the maximum spread of the walker for up to a specific time-step (say, $t=15$). This site-dependent set remains constant throughout the specified time steps, representing one disorder realization. In our study, any CQW measure $\Omega_1(t)$, such as the entanglement entropy ($E_{av}$) or the walker's position probability ($P_{av}(x=0)$), is first evaluated for a specific disorder realization. In this study, 500 such disorder realizations are computed to estimate the realization-average (quenched average, as one random-phase set is fixed to a particular realization)~\cite{sasha}, $\langle \Omega_1(t) \rangle$, \newline i.e.,
$\langle \Omega_1(t) \rangle=\frac{1}{500}\sum_{j=1}^{500}\Omega_1 (\ket{\psi(t)}_j\bra{\psi(t)}_j).\;\;$
\newline Static phase disorder does not affect the hopping length (i.e., $J(t)=1$, $\forall t$) nor the qubit coin, $\hat{H}$. Thus, the \textit{odd-even} parity is preserved as in the clean-CQW case for 4-cycle. It is shown in SM Fig.~2(a), where we see the probability of finding the walker $\langle P_{av}(x= 0)\rangle$ at odd time steps is always zero. Phase disorder also does not hinder the walker's \textit{physical time-cone} for the same reason as also seen for the clean-CQW.

Interestingly, SPE dynamics shows significant differences under static phase disorder for different disorder strengths, i.e., for $\delta$ ranging from 0 (clean-CQW) to 1 in units of $\pi$, see Fig.~\ref{f3} for 4-cycle. From Fig.~\ref{f3}(a-b) and SM Fig.~3 for odd 3-cycle, we observe that MESPS, at $t=1$, is robust against static phase disorder of any arbitrary strength. To prove this robustness, we find the time-evolved state, with Eq.~(\ref{eq1}) for $\phi=\frac{\pi}{2}$,
\begin{equation}
\ket{\psi(t=1)}=\hat{S}_{sp}\cdot[\sum_{x}\ket{x}\bra{x}\otimes \hat{H}]\frac{\cos(\frac{\theta}{2})\ket{0_p,0_c}+i\sin(\frac{\theta}{2})\ket{0_p,1_c}}{\sqrt{2}}.
\label{eq22n}
\end{equation}
Using Eq.~(\ref{eq8}), one obtains from Eq.~(\ref{eq22n}),
\begin{equation}
\begin{split}
\ket{\psi(t=1)}&=\frac{1}{\sqrt{2}}[(\cos(\frac{\theta}{2})+i\sin(\frac{\theta}{2}))e^{i\alpha_0(0)} \ket{(k-1)_p,0_c}\\&\;\;\;\;\;+(\cos(\frac{\theta}{2})-i\sin(\frac{\theta}{2})) \ket{1_p,0_c}],
\end{split}
\label{eq24n}
\end{equation}
and from its density matrix $\ket{\psi(t=1)}\bra{\psi(t=1)}$, we get the reduced density matrix,
\newline$
\rho_c(t=1)=\text{Tr}_p(\ket{\psi(1)}\bra{\psi(1)}), \text{ i.e., } \rho_c(t=1)=\frac{1}{2}\begin{pmatrix}
1 & 0\\
0 & 1
\end{pmatrix},$ which is maximally mixed independent of the choice of initial state parameter $\theta$, random phase angles $\alpha_0(0)$ and the number of sites ($k$) in the cyclic graph. The entanglement entropy of this state is,
\begin{equation}
E(\rho_c)=-\text{Tr}(\rho_c\log_{2}\rho_c)=\frac{1}{2}\log_2(2)+\frac{1}{2}\log_2(2)=1,
\label{eq25}
\end{equation}
which proves the state in Eq.~(\ref{eq24n}) has maximal SPE and is a MESPS (a single-particle Bell state). The MESPS at $t=1$ is robust against static phase-disorder of arbitrary strength and for any initial state, i.e., for an infinite number of initial states (Eq.~(\ref{eq1})) with arbitrary $\theta$; also see Fig.~\ref{f3}(b). The recurrence or periodicity of SPE via 4-cycle CQW is lost for phase disorder, and no recurrence occurs for 3-cycle under such disorder, i.e., it remains chaotic even under such disorder, also see SM Fig.~3. From Fig.~\ref{f3}(b), one can observe that SPE remains almost unaffected by small disorder strengths $\delta$; see, for example, SPE values at $t=3,5,7,...$. These results are valid for small time-steps ($t \le 8$), see Fig.~\ref{f3}(a)-(b); for larger $t$, SPE is vulnerable to all disorder strengths due to the strong coupling between the evolved state and disorder, i.e., the disorder effect piles up in course of time and at large $t$, disorder effects on SPE can be quite significant. Further, SPE increases under the phase disorder too, see for example SPE values at $t=6,10,14,...$, in Fig.~\ref{f3}(a)-(b). 
In addition, we observe that the static phase disorder can revive single-particle entanglement from zero, i.e., from an otherwise separable state observed at $t=4,8,12$, see  Fig.~\ref{f3}(b) (cyan, yellow, grey) and Fig.~\ref{f3}(a) (blue). We find similar results for the 3-cycle case, see  SM (Fig.~3), i.e., SPE can increase or decrease for static phase disorder of appropriate strength. Moreover, at $t=1$, MESPS is resilient to static phase disorder and initial state choice, as seen in the 4-cycle case.

\textcolor{brown}{\textit{Dynamic-phase disorder.--}}
The dynamic-phase disorder can be introduced in the CQW setup through the following translation operator (from Eq.~(\ref{eq4})), i.e.,
$\hat{S}_{dp}(t) = \sum_{s=0}^{1}\sum_{x=0}^{k-1}\ket{((x+2s-1) \text{ mod } k)_p}\bra{x}\;e^{i\beta_s(t)}\otimes\ket{s_c}\bra{s_c},
$
where the time-dependent relative phase angles $\beta_0(t)$ (with keeping $\beta_1(t)=0$, for the similar reason discussed below Eq.~(\ref{eq8})) are random numbers $[0,\delta]$ in units of $\pi$ drawn from uniform distribution, i.e., Eq.~(\ref{eq9}) with $\beta_0(t)$ in the place of $\alpha_0(x)$. The full CQW evolution is now, $U_{k}(t) = \hat{S}_{dp}(t)\cdot[\sum_{x}\ket{x}\bra{x}\otimes \hat{H}]\;$ (which is different from $U_{k}(t)$ for static-phase disorder). Finally,
\begin{equation}
\ket{\psi(t)}=\hat{S}_{dp}(t)\cdot[\sum_{x}\ket{x}\bra{x}\otimes \hat{H}]\ket{\psi(t-1)}\;.
\end{equation}

In the numerical simulation of the dynamic disorder, unlike the static-phase case, a set of time-dependent (and site-independent) random phase $\beta_0(t)$ angles is generated, with its length equal to the maximum time steps considered for the CQW dynamics (say, $t=15$). It represents one disorder realization, and subsequently, a realization average is computed with 500 such realizations for any quantity $\Omega_2(t)$, such as $E_{av}$ or $P_{av}(x=0)$, \newline i.e.,
$
\langle \Omega_2(t) \rangle=\frac{1}{500}\sum_{j=1}^{500}\Omega_2 (\ket{\psi(t)}_j\bra{\psi(t)}_j)\;.
$
A dynamic phase disorder preserves both the \textit{odd-even} parity for a 4-cycle and the particle's \textit{physical} \textit{time-cone} for all cycles, as also seen for the static variant; see SM Fig.~2 (red).
\begin{figure}[h!]
\includegraphics[width=9.8cm,height=4.5cm]{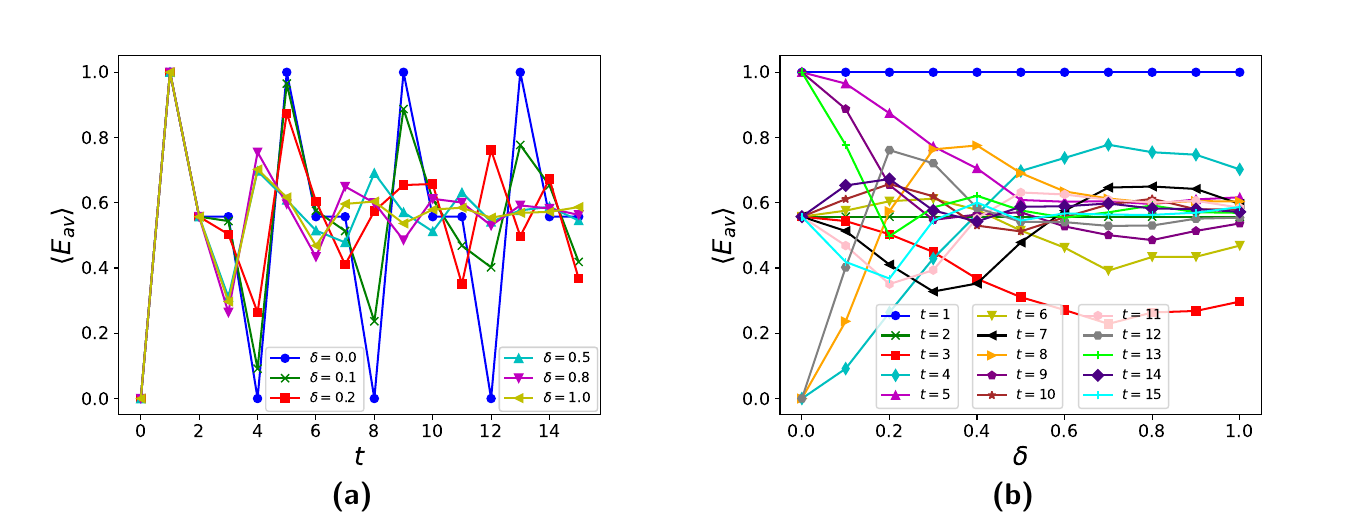}
\caption{(a) SPE $\langle E_{av} \rangle$ vs. time step $t$ for different dynamic-phase-disorder strength ($\delta$); (b) SPE $\langle E_{av} \rangle$ vs. dynamic-phase-disorder strength $\delta$ for different $t$ for 4-cycle.}
\label{f5}
\end{figure}

Interestingly, the SPE dynamics shows remarkable divergence under the dynamic-phase disorder of different strengths, i.e., for $\delta$ ranging from 0 (clean-CQW) to 1 in units of $\pi$, see Fig.~\ref{f5} and SM Fig.~4. SPE (MESPS) recurrence via 4-cycle CQW is lost for dynamic
phase disorder, and no recurrence occurs for 3-cycle. From Fig.~\ref{f5}(a-b), we observe that MESPS, at $t=1$, is robust against dynamic phase disorder of arbitrary strength too, i.e., it gives a Bell (MESPS) state irrespective of the phase disorder strength, also see SM Sec.~B for its detailed proof. This result holds for any initial state, see Eq.~(\ref{eq1}) and for any arbitrary $\theta$ (with $\phi=\frac{\pi}{2}$). From Fig.~\ref{f5}(b), one can observe that SPE is almost unaffected for small disorder strengths $\delta$, see SPE values at $t=3,5,7,...$, and SPE increases due to dynamic-phase-disorder, see SPE values at $t=4,6,8,10,12,14,...$ In addition, we observe that the dynamic phase disorder revives single-particle entanglement from zero, i.e., from an otherwise separable state, see Fig.~\ref{f5}(a) (blue) and Fig.~\ref{f5}(b) (cyan, yellow, grey) at $t=4,8,12$. We find similar results for the 3-cycle case, see SM  Fig.~4, i.e., SPE can increase and decrease due to dynamic phase disorder. Moreover, for the 3-cycle, at $t=1$, MESPS is resilient to dynamic-phase disorder and initial state choice, as also seen for the 4-cycle case, see SM Sec.~B.

\textcolor{brown}{\textit{Static vs dynamic phase disorder.--}}
Although static and dynamic phase disorders have similar effects on the CQW and generated SPE values, like preserving odd-even parity (4-cycle), obeying physical time-cone and increasing SPE as well as reviving SPE from zero, static-phase disorder, on average, is the stronger phase-disorder, see SPE values at $t=4,8,12,15$ in Figs.~\ref{f3}-\ref{f5}. There, we see that static phase disorder reduces SPE more than the dynamic variant.

\textcolor{brown}{\textit{Coin disorder and SPE in CQW.--}} Coin disorder in CQW setup can also be site-dependent (static) or time-dependent (dynamic). The former is introduced in the CQW setup through modified evolution, i.e., $U_{k}(t) = \hat{S}(t)\cdot[\sum_{x}\ket{x}\bra{x}\otimes \hat{C}(\rho(x))]\;$ with site-dependent coin operators~\cite{sasha} and the latter via, $U_{k}(t) = \hat{S}(t)\cdot[\mathbf{I}_P\otimes \hat{C}(\rho(t))]\;$, with time-dependent coin operators. For both static and dynamic coin disorder of strength ($\omega$), translation operator is $\hat{S}(t)$ (see, Eq.~\ref{eq4}), with hopping length $J(t)=1, \forall t$, and MESPS at $t=1$ remains unperturbed for both 3- and 4-cycles, for two phase-symmetric initial states, i.e., $\frac{\ket{0_p,0_c}\pm i\ket{0_p,1_c}}{\sqrt{2}}$, see Eq.~(\ref{eq1}) with $\theta=\frac{\pi}{2}$ with $\phi=\frac{\pi}{2},\frac{3\pi}{2}$. See SM Sec.~C for its complete analytical derivation and more details. Therein, we also observe that SPE remains almost unaffected by low levels of coin disorder. The static and dynamic coin disorders can also increase SPE at specific time steps. In addition, we observe that both the static and dynamic coin disorders revive SPE from zero, i.e., from an otherwise separable state observed at $t=4,8,12$, see SM Sec.~C and Fig.~5(b). Static and dynamic coin disorder have many similar effects on CQW and generated SPE values, such as both preserving odd-even parity (4-cycle), obeying physical time-cone (all cycles), increasing SPE (at specific time steps $t$) as well as reviving SPE from zero. However, static coin disorder, on average, is the more substantial coin disorder, see SM Sec.~C and Figs.~5-8, where we see the static coin disorder reduces SPE more than the dynamic variant for both 3- and 4-cycles. 


\textcolor{brown}{\textit{Position disorder and SPE in CQW.--}}
Position disorder can be introduced in CQW dynamics via the evolution, $U_{k}(t) = \hat{S}(t)\cdot[\mathbf{I}_P\otimes \hat{H}]\;$, where $\mathbf{I}_P$ is the identity operator in $\mathcal{H}_P$ space. In this case, the time-dependent translation operator $\hat{S}(t)$ is given by Eq.~(\ref{eq4}) with the hopping length $J(t)$ at any $t$, drawn from a Poisson distribution (PD) of different mean values $\lambda$, i.e., $\frac{\lambda^b e^{-\lambda}}{b!}, \text{with $b$ being the event count in PD~\cite{aditi19}.} $ Therefore, in this case, $\ket{\psi(t)}=\hat{S}(t)\cdot[I_P\otimes \hat{H}]\ket{\psi(t-1)}$. Additionally, we generate an array of time-dependent random $J(t)$ values for the numerical simulation, with the array length equal to the maximum time steps considered in the CQW dynamics (say, $t=15$). It represents one disorder realization, and subsequently, the realization average is computed with 500 such realizations for estimating the quantity $\Omega_5(t)$ such as entanglement entropy $E_{av}$ or $P_{av}(x=0)$, which is defined as,
$
\langle \Omega_5(t) \rangle=\frac{1}{500}\sum_{j=1}^{500}\Omega_5 (\ket{\psi(t)}_j\bra{\psi(t)}_j)\;.\;
$
\begin{figure}[h!]
\includegraphics[width=9.8cm,height=4.5cm]{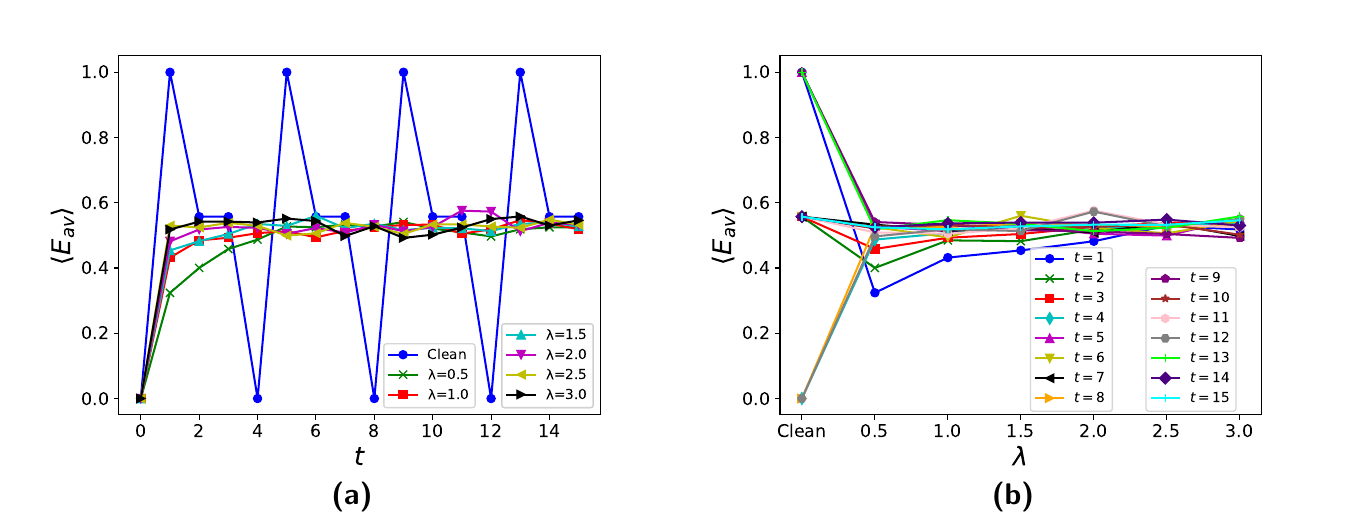}
\caption{(a) SPE $\langle E_{av} \rangle$ vs. time step $t$ for different position-disorder strength ($\lambda$); (b) SPE $\langle E_{av} \rangle$ vs. position-disorder strength $\lambda$ for different $t$ for 4-cycle.}
\label{f11}
\end{figure}

As the hopping (jump) length is a random number, the position disorder violates \textit{odd-even} parity, which was obeyed in clean CQWs and also in coin and phase disordered CQW, see SM Fig.~2(a) (yellow curve). We observe that $\langle P_{av}(x=0)\rangle \neq 0$ at odd $t$, confirming the violation of the \textit{odd-even} parity for 4-cycle CQW evolution. It begets a rich class of superposed and entangled quantum states with finite probabilities of finding the particle at most sites. It helps design cryptographic keys~\cite{p2,p4}, as a highly probabilistic (non-inferable) public key correlates with greater security in a quantum cryptographic algorithm than a deterministic public key~\cite{p2,p4,pqc}. Furthermore, the position disorder breaks the physical \textit{time cone }of the walker and thus can enhance the particle spreading, which helps design search algorithms and study transport phenomena~\cite{eqw1,eqw2,naves22,portugal}.

However, SPE is more vulnerable to position disorder than phase or coin disorder. The recurrence or periodicity of the SPE (or MESPS of clean-CQW) in 4-cycle CQW is distorted under position disorder; see Fig.~\ref{f11} above and SM Fig.~2. Moreover, under position disorder, SPE saturates to a fixed value in both 4-cycle and 3-cycle, irrespective of the disorder strength $\lambda$, at large $t$; see Fig.~\ref{f11} and SM Sec.~D (Fig.~10).
Interestingly, in line with other disorders, a position disorder can also revive SPE at time steps where no SPE was generated via the clean-QW with any arbitrary initial state (see Fig.~\ref{f11} above).

\textcolor{brown}{\textit{Analysis.--}}
The effects of phase, coin, and position disorder on CQW evolution and SPE are juxtaposed in Table.~\ref{tab}. MESPS at $t=1$ is robust against phase disorder (static and dynamic) of arbitrary strength irrespective of the choice of the initial state. On the other hand, against coin disorder (static and dynamic) of arbitrary strength, MESPS at $t=1$ is robust for only two phase-symmetric initial states. Also, the reduction in SPE (at any $t$) is insignificant to low levels of phase and coin disorders (static and dynamic). All disorders are shown to revive SPE from zero and enhance SPE at specific time steps. Unlike coin and phase disorders, position disorder causes SPE to converge to a fixed value at large t (irrespective of the disorder strength and initial state choice). An algorithm to see the effects of disorders via our CQW setup on SPE is provided in SM Sec.~E.
\begin{table}[h!]
\centering
\resizebox{\columnwidth}{!}{ 
\begin{tabular}{|c|c|c|c|c|c|}
\cline{1-6}
\textbf{Property\textcolor{blue}{$\downarrow$} / Disorder\textcolor{blue}{$\rightarrow$}} & \multicolumn{2}{c|}{\textbf{Static}} & \multicolumn{2}{c|}{\textbf{Dynamic}} & \textbf{Position} \\
\cline{2-5}
& \textbf{Phase} & \textbf{Coin} & \textbf{Phase} & \textbf{Coin} &  \\
\hline
\begin{tabular}[c]{@{}p{3.2cm}@{}} Robust MESPS (Bell state) at  $t=1$ to \\disorder\end{tabular} & \begin{tabular}[c]{@{}p{2.3cm}@{}} Yes, for all initial states and disorder strengths\\(see Eqs.~(\ref{eq24n}-\ref{eq25}))\\ \end{tabular} & \begin{tabular}[c]{@{}p{2.3cm}@{}} Yes, for phase symmetric initial states and all disorder strengths \\(see SM Eqs.~(12-16)) \end{tabular}& \begin{tabular}[c]{@{}p{2cm}@{}} Yes, for all initial states and disorder strengths\\(see SM Eq.~(7))\\ \end{tabular} &\begin{tabular}[c]{@{}p{2.3cm}@{}} Yes, for phase symmetric initial states and all disorder strengths\\(see SM Eqs.~(17-19)) \end{tabular} & No\\
\hline

\begin{tabular}[c]{@{}p{3.2cm}@{}} Resilience of SPE to small disorder \end{tabular} & Yes & Yes &Yes & Yes & No\\
\hline

\begin{tabular}[c]{@{}p{3.2cm}@{}}Revival of SPE from zero\end{tabular}& Yes& Yes & Yes& Yes& Yes\\
\hline
\begin{tabular}[c]{@{}p{3.2cm}@{}}Enhancement of SPE\end{tabular}& Yes& Yes & Yes& Yes& Yes\\
\hline
\begin{tabular}[c]{@{}p{3.2cm}@{}} Convergence of SPE to a fixed value \end{tabular} & No & No& No& No& \begin{tabular}[c]{@{}p{1.7cm}@{}} Yes, for large disorder strengths\end{tabular} \\
\hline
\end{tabular}

}
\caption{Disorder effects on SPE in CQW.}
\label{tab}
\end{table}
We have also compared our results with prior works, see Refs.~\cite{sasha,prl,china,ntuyong,naves22} on DTQW with 1D line in SM Table~I and SM Sec.~F. Refs.~\cite{sasha,prl,china,ntuyong,naves22} focus on generating SPE using disorder (these works considered either a coin or phase or position disorder) and that too for some particular initial states. These proposals do not look at the effects of a disorder on MESPS or SPE (via clean QW). We bridge this gap and comprehensively report on generating SPE via disorder. We also discuss the effects of various disorders on SPE's enhancement, revival, and resilience under phase, coin, and position disorder of arbitrary strength with a most general initial state. 

This proposal can be experimentally realized using single-photons with optical elements (e.g., polarizing beam splitters, waveplates, Jones plates),  wherein the photon’s
polarization encodes the coin state and time bins or orbital angular momentum encodes the position state~\cite{bian,2dqw-expt,jplate,expt1-science, Chandra2022}. 

\textcolor{brown}{\textit{Conclusion and outlook.--}}
This study explores the robustness, possibility of revival, and vulnerability of single-particle entanglement (SPE) under phase, coin, and position disorder. Disorders destroy the recurrence of SPE in CQW. The resilience of SPE to minor disorders, particularly phase and coin disorder, demonstrates its potential for stable large-scale QI processing. Phase disorder, especially, leaves maximal SPE (MESPS) states (at $t$=1) completely unperturbed for any disorder strength, irrespective of the choice of the initial state, i.e., for any arbitrary $\theta$ in Eq.~(1). Further, phase-symmetric initial states are entirely immune to coin disorder of arbitrary strength. Moreover, phase and coin disorder enhances SPE at specific time steps. These also revive SPE from zero, i.e., from an otherwise separable state, irrespective of the choice of the initial state.

However, unlike coin or phase disorder, position disorder is more deleterious, breaking the \textit{odd-even} parity and distorting the walker's physical \textit{time cone}. Position disorder thus generates rich quantum states, which are helpful in quantum cryptography in designing cryptographic keys, as these states will be more challenging to decipher by eavesdroppers as compared to trivial states (product or symmetrically superposed states)~\cite{p2,p4,pqc}. Besides, position disorder can enhance particle dispersion, increase position standard deviation, and be a resource for an efficient quantum search algorithm and in analyzing transport phenomena~\cite{eqw1,eqw2,naves22,portugal}. Interestingly, despite its disruptive nature, position disorder causes SPE to converge to a fixed value at large $t$, indicating a level of robustness of SPE even under significant disturbances. These findings underline the nuanced role disorder plays in the evolution of SPE, offering valuable insights for optimizing quantum communication, dynamics of intricate quantum systems in a lab setting, and cryptographic protocols that leverage single-particle entangled states.

\newpage

\vspace{10cm}

\onecolumngrid
\section*{\underline{Supplementary Material} for "Phase vs coin vs position disorder as a probe for the resilience and revival of single particle entanglement in cyclic quantum walks"}
\centerline{Dinesh Kumar Panda$^{1,2}$,\; Colin Benjamin$^{1,2}$}

\centerline{$^{1}$School of Physical Sciences, National Institute of Science Education and Research Bhubaneswar, Jatni 752050, India}
\centerline{$^{2}$Homi Bhabha National Institute, Training School Complex, Anushaktinagar, Mumbai
400094, India}

\vspace{1cm}

\onecolumngrid
Herein, we discuss in Sec.~\textcolor{blue}{A} the generalized evolution dynamics that yield the quNit-qubit SPE via our CQW (cyclic quantum walk) setup encompassing the most general initial state, general shift, and an arbitrary coin. Then, we detail how Hadamard coin results in recurrent MESPS irrespective of the initial state, a particular case of generalized CQW dynamics. Then we discuss the properties of clean-CQW (generalized Hadamard walk): \textit{odd-even} parity, physical \textit{time cone}, recurrent MESPS. Therein, we also detail the effects of phase, coin, and position disorder on the \textit{odd-even} parity and physical \textit{time cone}. In Sec.~\textcolor{blue}{B}, we proved the robustness of MESPS at time-step $t=1$ to dynamic phase disorder of arbitrary strength irrespective of the choice of the initial state for both 3- and 4-cycles. Then, we discuss the effects of the static and dynamic phase disorders on SPE generated via CQW on a 3-cycle. Furthermore, Sec.~\textcolor{blue}{C} discussed the implementation of static and dynamic coin disorders in our CQW setup for 3- and 4-cycles and their effects on CQW evolution and generated SPE. Therein, we also discuss the robustness of MESPS at $t=1$ against the static and dynamic coin disorders for two phase-symmetric initial states in both 3- and 4-cycles, which we prove rigorously. Sec.~\textcolor{blue}{D} of this supplementary material discusses the effects of position disorder of varying strengths on SPE via CQW for 3-cycle. We provide an algorithm for implementing disordered CQW and disorder effects on SPE for interested researchers in Sec.~\textcolor{blue}{E}. Finally, in Sec.~\textcolor{blue}{F}, we compare our results with prior works~\cite{sasha,prl,china,ntuyong,naves22} on DTQW with 1D line which focuses on generating SPE using disorder (each work considered either a coin or phase or position disorder) and that too for some specific initial states.

\subsection{Generalised CQW dynamics, properties and SPE generation}

\subsubsection{Generalised CQW and SPE}
\begin{figure}[h!]
\includegraphics[width=12cm,height=3.5cm]{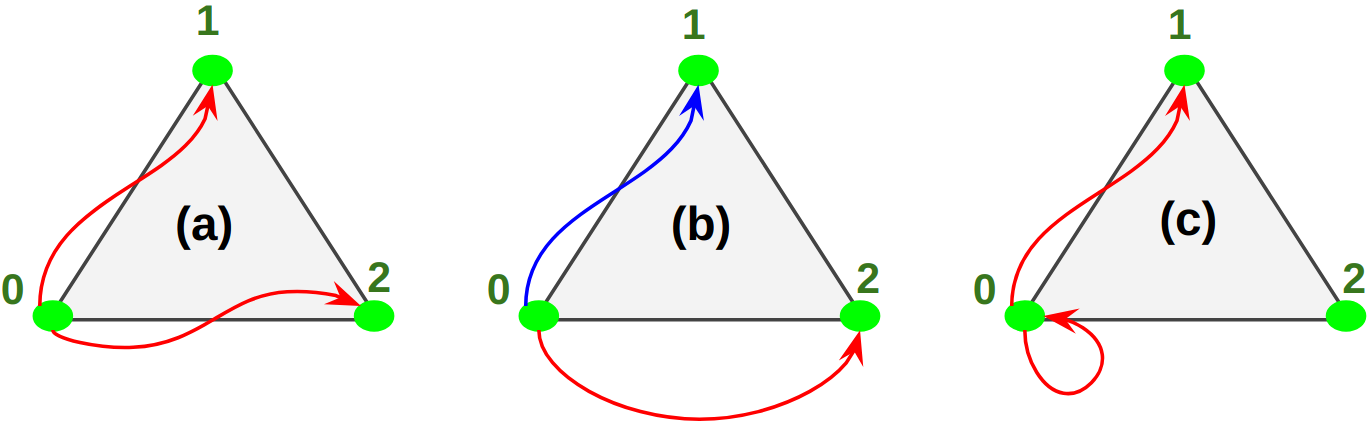}
\caption{(a) Clean-CQW (Hadamard walk) on a 3-cycle ; (b) CQW with phase or coin disorder on a 3-cycle, where amplitudes in both hopping change;(c) CQW with position disorder on a 3-cycle where the hopping length changes randomly. Sites are shown as green dots, and red-blue curves denote the hopping of the particle on a 3-cycle.}
\label{f1b}
\end{figure}

The cyclic quantum walk (CQW) on a $k$-cycle graph (e.g., 3-cycle, see Fig.~\ref{f1b} and 4-cycle, see main text Fig.~1), see main text Eq.~(2), is utilized to find the following recursion relation for the time evolution of the quantum walker (particle),
\begin{align}
\begin{split}
\ket{\psi(t+1)} =&\hat{S}.[\sum_{x}\ket{x_p}\bra{x_p}\otimes \hat{C}(\rho)]\ket{\psi(t)},
\\
=&\sum_{x=0}^{k-1}[\alpha_{0}(x-J',t+1)\ket{((x-J')\text{ mod } k)_p,0_c}+ \alpha_{1}(x+J',t+1)\ket{((x+J')\text{ mod } k)_p,1_c}],
\end{split}
\label{eq6}
\end{align}
where,
$\alpha_{0}(x-J',t+1)=\sqrt{\rho}\;\alpha_{0}(x,t)+\sqrt{1-\rho}\;\alpha_{1}(x,t)$,\;\;\;$\alpha_{1}(x+J',t+1)=\sqrt{1-\rho}\;\alpha_{0}(x,t)-\sqrt{\rho}\;\alpha_{1}(x,t)$.
Moreover, $J'$ is the hopping length at time-step $t+1$. Eq.~(\ref{eq6}) indicates that the evolved quantum state exhibits high-dimensional (quNit-qubit) single particle entanglement between the position (quNit, $N=k$, which is a $k$-level system) and the coin (qubit) degrees of freedom of the walker, as it is no longer a product state.

For instance, for obtaining SPE from CQW dynamics and to check the applicability of the recursion relation (\ref{eq6}), let us set at $t=0$, $\phi=\frac{\pi}{2}$ and consider arbitrary $\theta$ (i.e., an arbitrary initial state, see main text Eq.~(1)). We have $\ket{\psi(t=0)} =\cos(\frac{\theta}{2})\ket{0,0_c} + i\sin(\frac{\theta}{2})\ket{0,1_c}$, with amplitudes, $\alpha_{0}(0,0)=\cos(\frac{\theta}{2})$ and $\alpha_{1}(0,0)= i\sin(\frac{\theta}{2})$. Now using the  Eq.~(\ref{eq6}) with $J'=1$, we find $\alpha_{0}(k-1,1)=\frac{1}{\sqrt{2}}(\cos(\frac{\theta}{2})+i\sin(\frac{\theta}{2}))$ and  $\alpha_{0}(1,1)=\frac{1}{\sqrt{2}}(\cos(\frac{\theta}{2})-i\sin(\frac{\theta}{2}))$ and the state at $t=1$ is,

\begin{equation}
\ket{\psi(t=1)}= \alpha_{0}(k-1,1)\ket{(k-1)_p,0_c}+\alpha_{0}(1,1)\ket{1_p,1_c}.
\label{bell}
\end{equation}
In particular, for $k=4$ (4-cycle), we get $\ket{\psi(t=1)}= \alpha_{0}(3,1)\ket{3_p,0_c}+\alpha_{0}(1,1)\ket{1_p,1_c}$ and for $k=3$ (3-cycle), $\ket{\psi(t=1)}= \alpha_{0}(2,1)\ket{2_p,0_c}+\alpha_{0}(1,1)\ket{1_p,1_c}$.
For Hadamard coin $\hat{H}$ (i.e., $\rho=\frac{1}{2}$), this evolved state $\ket{\psi(t=1)}$ is a single-particle Bell state (i.e., a MESPS) and possesses maximal SPE for any value of $\theta$ as numerically also shown in Ref.~\cite{p2}. We can see that this result is independent of the number of sites ($k$) on the cyclic graph. As the CQW evolution with $\hat{H}$ coin yields ordered or periodic CQW dynamics on a 4-cycle, therefore, this yields recurrent maximal SPE or maximally entangled single-particle states (MESPS) with period 4, i.e., at $t=1,5,9,13,...$, see Ref.~\cite{p2}, and see~Fig.~2(a) (blue) in main text.
\subsubsection{Properties of clean-CQW: odd-even parity, physical time cone, recurrent MESPS}
\begin{figure}[h!]
\includegraphics[width=18cm,height=8.3cm]{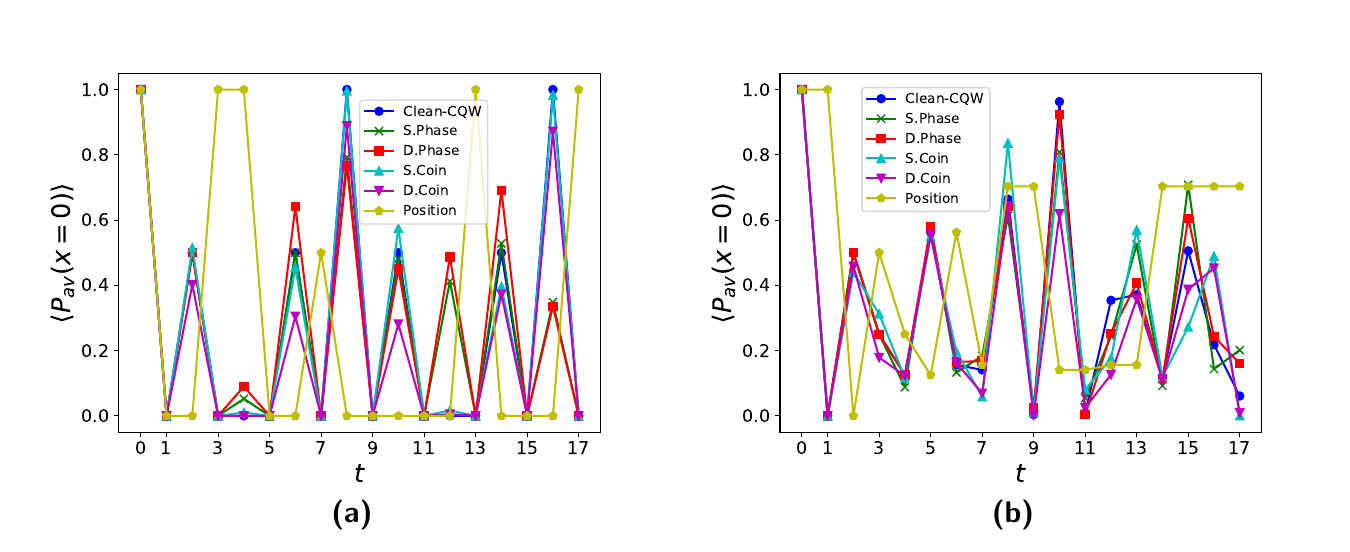}
\caption{Probability $\langle P_{av}(x=0) \rangle$ vs. time step $t$ in clean-CQW (blue), CQW with static phase (green), dynamic phase (red), static coin (cyan), dynamic coin (purple) and position (yellow) disorders in 4-cycle (a) and 3-cycle (b). Odd-even parity is broken in 4-cycle only for position disorder (preserved in clean-CQW or CQW with phase or coin disorders), whereas the same is broken in 3-cycle in the presence or absence of any disorder.}
\label{f2}
\end{figure}

As mentioned in the main text, we observe that a generalized Hadamard walk dynamics for an arbitrary general initial state (i.e., main text Eq.~(1)), referred to as clean-CQW (i.e., without disorder), has three fundamental properties:

1).~It preserves the \textit{odd-even} parity~\cite{sasha} for even $k$-cycle graphs, i.e., at even (odd) $t$, only even (odd)-numbered sites have finite probabilities of occupation for the quantum walker.
If $P(x_p=0_p)$ defines the probability of finding the walker at the origin (initial site, $\ket{x_p}=\ket{0_p}$) at time-step $t$, i.e.,
\begin{align}
\begin{split}
P(0_p) = |\bra{0_p,0_c}\ket{\psi(t)}|^2+|\bra{0_p,1_c}\ket{\psi(t)}|^2,
\end{split}
\label{eq5}
\end{align}
the \textit{odd-even} parity implies $P(0_p)=0$ at all odd time-steps (odd $t$). This parity is always preserved in even-site cycles such as 4-cycle, as shown in Fig.~\ref{f2}(a) (blue). Note that $P_{av}(x=0) = \frac{1}{\pi} \int_{0}^{\pi} P(0_p)\;d\theta\; $ and Fig.~\ref{f2} results hold for any $\theta$ value, i.e., any arbitrary initial state (main text Eq.~(1) with $\phi=\frac{\pi}{2}$). However, for odd cycle graphs like 3-cycle, it is not preserved as the dynamics can be chaotic\cite{p2}, see Fig.~\ref{f2}(b) (blue). Physically, the \textit{odd-even} parity preservation (in even $k$-cycle graphs) is a consequence of the constant hopping length (e.g., $J(t)=1$, $\forall t$) and Hadamard coin operator creates an equal superposition of the coin basis states at any arbitrary time step; see main text Fig.~1(a). In other words, detectors can not detect the photon (a physical walker) at even sites once the photon has evolved to an odd time step (i.e., if $t$ is odd).

2).~It respects the \textit{physical time-cone} of the walker, i.e., a walker at position site $\ket{x_p}=\ket{m_p}$ at time step $t$ will have finite probability of being found at positions up to $\ket{x'_p}=\ket{((m\pm n)\text{ mod } k)_p}$  at the next $n^{\text{th}}$ time-step ($t+n$), and the particle (photon) is not to be found anywhere else beyond $\ket{x'_p}$. It is guaranteed by $J(t)=1$, independent of the time steps or sites, and it does not allow the spread of the walker in the physical position-Hilbert space beyond one hopping unit. This universal property applies to both odd and even $k$-cycles.

3).~It generates MESPS at time step $t=1$ for any arbitrary $k$-cycle graph. Corollary: for 4 and 8-cycles, MESPS is generated recurrently with period 4 at time step $t=1,5,9,...$, see main text Fig.~2(a) (blue) and also~\cite{p2}. The generic time-evolution of CQW dynamics with the general initial state (main text Eq.~(1)) with arbitrary qubit-coin (main text Eq.~(3)) and general translation operator (main text Eq.~(4)), which generates high-dimensional (quNit-qubit) single particle entanglement between the position (quNit, $N=k$) and the coin (qubit) DoF of the walker, is detailed in Sec.~A.1 above.
\subsubsection{Features of disordered CQW: breaking of \textit{odd-even} parity \& violation of physical \textit{time cone}}
As shown in Fig.~\ref{f2}, static and dynamic phase disorders (e.g., with disorder strength $\delta=0.2$) obey odd-even parity. The same is observed in static and dynamic coin disorders (e.g., with disorder strength $\omega=0.2$). However, position disorder (e.g., with disorder strength $\lambda=1.5$) breaks the odd-even parity, i.e.,  $\langle P_{av}(x=0) \rangle \ne 0$ for odd time-step $t$. Moreover, position disorder breaks the physical time-cone of the walker, unlike coin or phase disorders. It is because, for position disorder, the hopping length $J(t)$ is a random number, unlike in the case of a clean-CQW or a coin/phase-disordered CQW.

Interestingly, breaking the \textit{odd-even }parity and the \textit{physical time-cone} of the walker has beneficial consequences for quantum-information-processing tasks and efficient algorithm design. The former generates a rich class of complex quantum states with a finite probability of finding the particle at all sites. It helps design quantum cryptographic keys~\cite{p2,p4}, as a highly probabilistic (non-inferable) public key correlates with
greater security in a cryptographic algorithm than a deterministic public key. On the other hand, the latter can enhance the particle spreading and increase the position standard deviation, which is a fundamental resource for an efficient search algorithm and transport phenomena~\cite{eqw1,eqw2,naves22,portugal}. At the same time, studying the behavior and resilience of generated coin-position SPE via Hadamard walk (clean-CQW) on $k$-cycles (with odd or even $k$) in the presence of disorder is crucial to understanding the fundamental evolution and entanglement dynamics of quantum systems in a lab setting, which is detailed in the main text.


\subsection{Phase disorder and SPE for 3-cycle and 4-cycle}
Herein, we prove the robustness of MESPS at $t=1$ for dynamic phase disorder for both 3-cycle and 4-cycle (proof of such robustness against static phase disorder is done in the main text). We then discuss the effects of static and dynamic phase disorders on SPE for the 3-cycle, supplementing the main text.
\subsubsection{Robustness of MESPS at $t=1$ for dynamic phase disorder for both 3 and 4-cycles}

\label{appb}
\begin{figure}[h!]
\includegraphics[width=18cm,height=8.5cm]{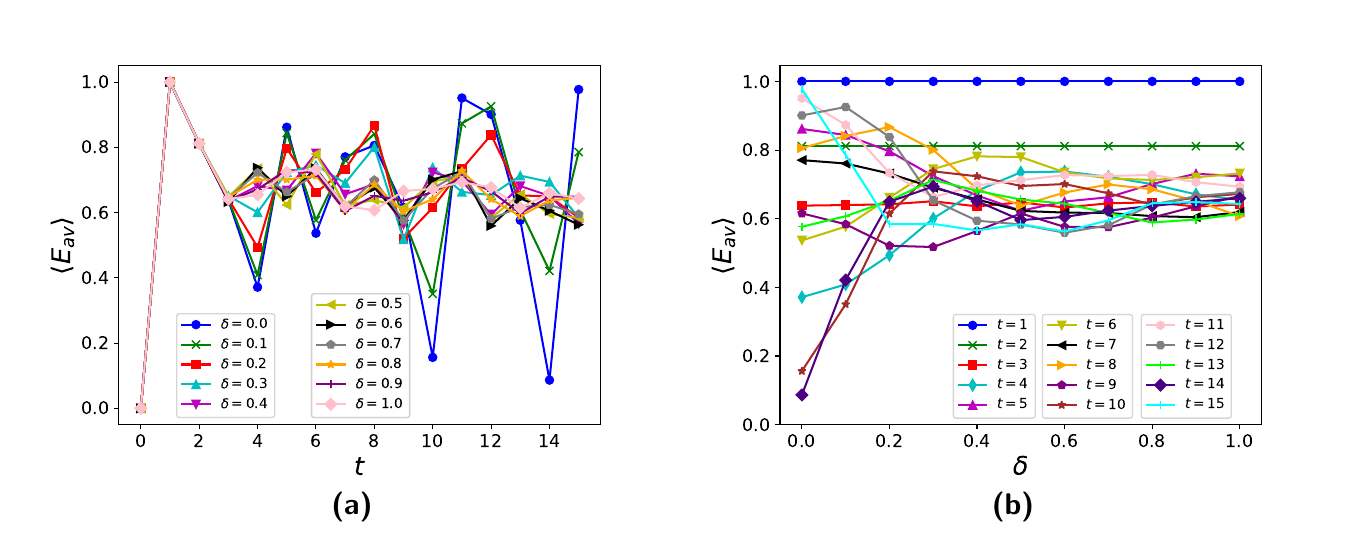}
\caption{(a) SPE $\langle E_{av} \rangle$ vs. time step $t$ for different static-phase-disorder strength ($\delta$); (b) SPE $\langle E_{av} \rangle$ vs. static-phase-disorder strength $\delta$ for different time-steps $t$ for 3-cycle.}
\label{f4}
\end{figure}

As proved in the main text Fig.~2 and Fig.~\ref{f4}, MESPS at $t=1$ is unperturbed against static phase disorder of arbitrary strength. This holds for any initial state (main text Eq.~(1)),  for arbitrary $\theta$ values, i.e., 
\begin{equation}
\ket{\psi(t=0)} =\ket{0_p}\otimes\ket{q_c}=\ket{0_p}\otimes(\cos(\frac{\theta}{2})\ket{0_c} + e^{i\phi}\sin(\frac{\theta}{2})\ket{1_c}),
\label{eq21}
\end{equation}
where $\theta \in [0, \pi]$ and the phase $\phi=\frac{\pi}{2}$.

 Proof for the full immunity of the MESPS at $t=1$ under dynamic phase disorder of arbitrary strength can be similarly derived for the static case. The evolved state at $t=1$ in this case reads, 
\begin{equation}
\ket{\psi(t=1)}=\hat{S}_{dp}(t=1)\cdot[\sum_{x}\ket{x}\bra{x}\otimes \hat{H}]\ket{\psi(0)},
\label{eq27}
\end{equation}
where $\hat{H}=\frac{1}{\sqrt{2}}\begin{pmatrix}
1 & 1\\
1 & -1
\end{pmatrix}$ is the Hadamard coin  and the translation operator introducing dynamic phase-disorder,
\begin{equation}
\hat{S}_{dp}(t=1)=\sum_{x=0}^{k-1}\{\ket{((x-1) \text{ mod } k)_p}\bra{x}\;e^{i\beta_0(t=1)}\otimes\ket{0_c}\bra{0_c}+\ket{((x+1) \text{ mod } k)_p}\bra{x}\otimes\ket{1_c}\bra{1_c}\}.
\label{}
\end{equation} 
Herein, we set $\beta_1(t=1)=0$, as mentioned in the main text on page 4. The evolved state Eq.~(\ref{eq27}) at $t=1$, is then,
\begin{equation}
\ket{\psi(t=1)}=\frac{1}{\sqrt{2}}[(\cos(\frac{\theta}{2})+i\sin(\frac{\theta}{2}))e^{i\beta_0(1)} \ket{(k-1)_p,0_c}+(\cos(\frac{\theta}{2})-i\sin(\frac{\theta}{2})) \ket{1_p,0_c}]. 
\label{eq29}
\end{equation} This also results in a maximally mixed reduced density matrix, $\rho_c(t=1)=\frac{1}{2}\begin{pmatrix}
1 & 0\\
0 & 1
\end{pmatrix},$ and yields entanglement entropy, 
$E(\rho_c)=1$. Thus, dynamic phase disorder does not affect MESPS at $t=1$ like the static-phase disorder. It holds for any disorder strength and any initial state (for arbitrary $\theta$ and $\phi=\frac{\pi}{2}$). 

\subsubsection{Phase disorder and SPE for 3-cycle}
\begin{figure}[h!]
\includegraphics[width=18cm,height=8.5cm]{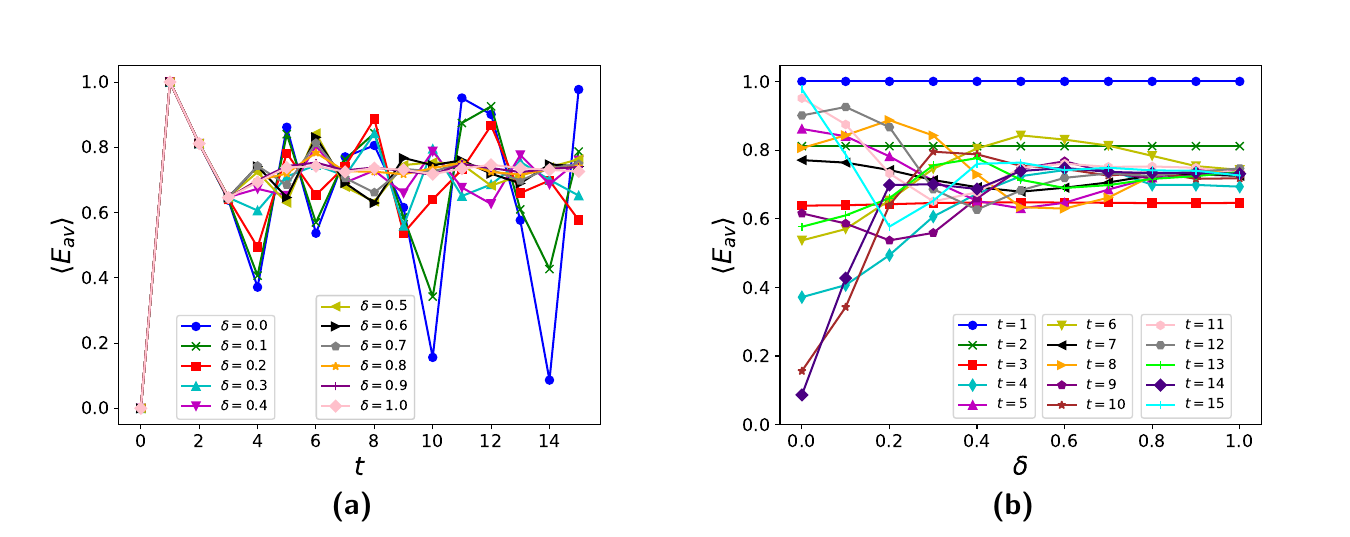}
\caption{(a) SPE $\langle E_{av} \rangle$ vs. time step $t$ for different dynamic-phase-disorder strength ($\delta$); (b) SPE $\langle E_{av} \rangle$ vs. dynamic-phase-disorder strength $\delta$ for different time-steps $t$ for 3-cycle.}
\label{f6}
\end{figure}

From Fig.~\ref{f4}, we observe that the static disorder can increase SPE at some time steps for (odd) 3-cycle (see Fig.~\ref{f1b}), e.g., see SPE at $t=4,6,10,13,14$. Similarly, from Fig.~\ref{f6}, we observe that the dynamic disorder can increase SPE at some time steps for odd 3-cycle, e.g., see SPE at $t=4,6,8,10,13,14$.

\subsection{Coin disorder and SPE for 3-cycle and 4-cycle}
 The main text mentions that coin disorder in the CQW setup can be site-dependent (static) or time-dependent (dynamic). Herein, we discuss the effects of static and dynamic coin disorders on the CQW evolution and the generated SPE.
\subsubsection{Static coin disorder}
The static-coin disorder can be introduced in the CQW setup through the following modified evolution, i.e., $U_{k}(t) = \hat{S}(t)\cdot[\sum_{x}\ket{x}\bra{x}\otimes \hat{C}(\rho(x))]\;$ with site-dependent coin operators~\cite{sasha} and $\hat{S}(t)$ is in main text Eq.~(4) with hopping length $J(t)=1, \forall t$. Here,
\begin{equation}
    \ket{\psi(t)}=\hat{S}(t)\cdot[\sum_{x}\ket{x}\bra{x}\otimes \hat{C}(\rho(x))]\ket{\psi(t-1)}\;.
\end{equation} The disorder strength is controlled by parameter $\omega$, i.e.,
$\rho(x)=\frac{1}{2}(1+\omega r_x)$, where $r_x\in[1,-1]$ are random numbers drawn from a uniform distribution~\cite{sasha}, i.e., main text Eq.~(6) with $r_x$ in the place of $\alpha_0(x)$ and $a=-1, b=1$. The clean-CQW (Hadamard walk) can be retrieved by setting $\omega=0$, 
and a fully disordered CQW is for $\omega=1$.
 In our numerical simulation of the static-coin disordered CQW, an array of random on-site  $r_x$ values is generated, with its size corresponding to the maximum spatial spread of the walker for up to a specific time-step (say, $t=15$). This site-dependent array remains constant throughout the specified time steps, representing one disorder realization. With 500 such disorder realizations, realization-average of any property $\Omega_3(t)$ for the CQW dynamics such as $E_{av}$ or $P_{av}(x=0)$, are evaluated as follows,
\begin{align}
\begin{split} 
\langle \Omega_3(t) \rangle=\frac{1}{500}\sum_{j=1}^{500}\Omega_3 (\ket{\psi(t)}_j\bra{\psi(t)}_j)\;.\;
\end{split}
\label{eq12}
\end{align}
Like the phase disorder, static coin disorder does not affect the hopping length (i.e., $J(t)=1$, $\forall t$), and the qubit-coin operator yields symmetric walker propagation. Thus, the \textit{odd-even} parity for 4-cycle and walker's \textit{physical time-cone} for all cycles are obeyed, as in the clean-CQW case, also see Fig.~\ref{f2}(cyan, purple) in Sec.~A.3 above. 

\begin{figure}[h!]
\includegraphics[width=18cm,height=8.5cm]{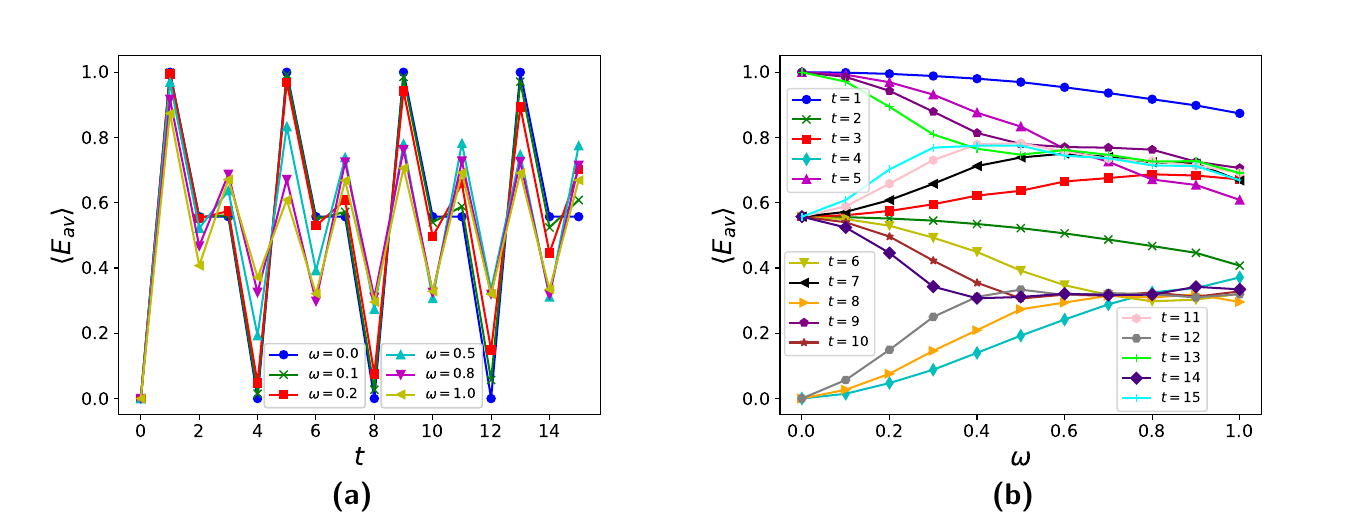}
\caption{(a) SPE $\langle E_{av} \rangle$ vs. time step $t$ for different static-coin-disorder strength ($\omega$); (b) SPE $\langle E_{av} \rangle$ vs. static-coin-disorder strength $\omega$ for different time-steps $t$ for 4-cycle.}
\label{f7}
\end{figure}
\begin{figure}[h!]
\includegraphics[width=18cm,height=8.5cm]{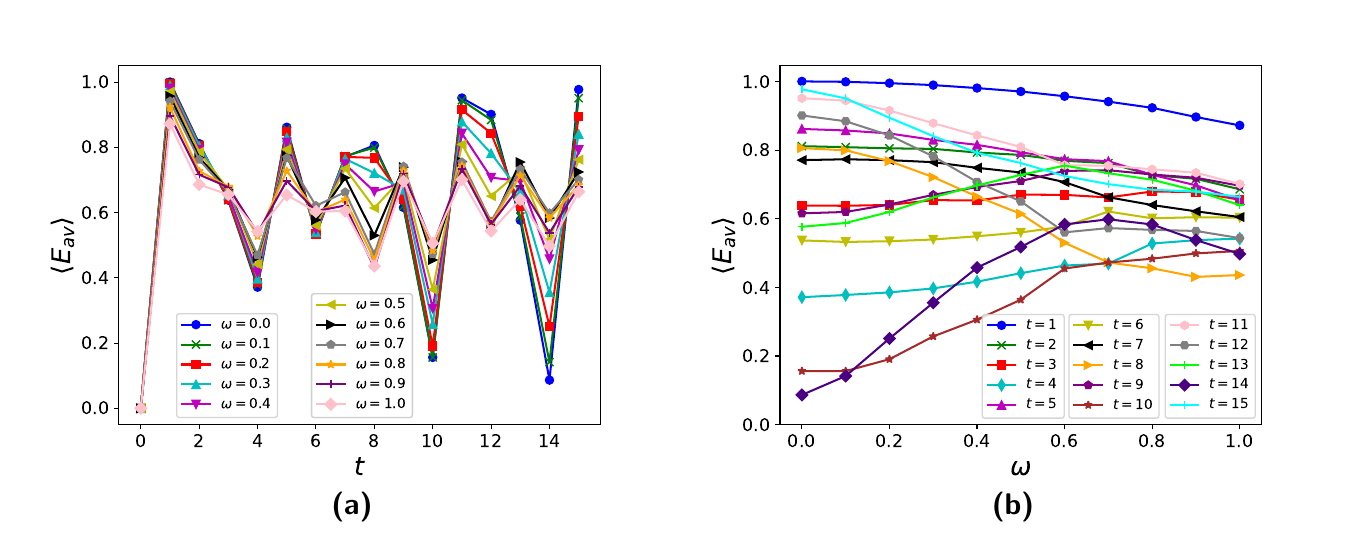}
\caption{(a) SPE $\langle E_{av} \rangle$ vs. time step $t$ for different static-coin-disorder strength ($\omega$); (b) SPE $\langle E_{av} \rangle$ vs. static-coin-disorder strength $\omega$ for different time-steps $t$ for 3-cycle.}
\label{f8}
\end{figure}

The quNit-qubit SPE dynamics show significant differences under static coin disorder for different disorder strengths, i.e., for $\omega$ ranging from 0 (clean-CQW) to 1, see Figs.~\ref{f7} and \ref{f8}. 
MESPS at $t=1$ remains unperturbed for both 3- and 4-cycles and any arbitrary disorder strengths, for two-phase symmetric initial states, i.e., Eq.~(1) with $\theta=\frac{\pi}{2}$ with $\phi=\frac{\pi}{2},\frac{3\pi}{2}$, i.e., $\frac{\ket{0_p,0_c}\pm i\ket{0_p,1_c}}{\sqrt{2}}$, also see below Subsec.~4 for its derivation. From Fig.~\ref{f7}(b), one can observe that SPE remains almost unaffected by small disorder strengths $\omega$; see, for example, SPE values at $t=1,3,5,7,9,...$ and the SPE increases under the static-coin disorder too, see for example SPE values at $t=3,4,7,8,11,12,15,...$. In addition, we observe that the static coin disorder revives single-particle entanglement from zero, i.e., from an otherwise separable state observed at $t=4,8,12$, see  Fig.~\ref{f7}(b) (cyan, yellow, grey). We find similar results for the 3-cycle case, see Fig.~\ref{f8}, i.e., SPE shows resilience or increase or decrease for static coin disorder of appropriate strengths. We find from Fig.~\ref{f8} that SPE is almost unaffected by static coin disorder of small strength. Further, we observe that the static coin disorder can increase SPE at some time steps for odd 3-cycle, e.g., see SPE at $t=4,6,9,10,13,14$, in  Fig.~\ref{f8}.

\subsubsection{Dynamic coin disorder }
To introduce a dynamic coin disorder, we consider the modified evolution operator, $U_{k}(t) = \hat{S}(t)\cdot[\mathbf{I}_P\otimes \hat{C}(\rho(t))]\;$, where $\mathbf{I}_P$ is the identity operator in $\mathcal{H}_P$ space, with hopping length $J(t)=1$ and time-dependent coin operators. Thus, \begin{equation}
    \ket{\psi(t)}=\hat{S}(t)\cdot[\mathbf{I}_P\otimes \hat{C}(\rho(t))]\;\ket{\psi(t-1)},
\end{equation} which is different than its static variant. Parameter $\omega\in[0,1]$, controls the disorder strength, i.e.,
$\rho(t)=\frac{1}{2}(1+\omega r_t)$, where $r_t\in[1,-1]$ are drawn from a uniform distribution of random numbers, see main text Eq.~(6) with $r_t$ in place of $\alpha_0(x)$ and $a=-1, b=1$. In the numerical simulation, unlike the static-coin disorder case, herein a set of time-dependent (and site-independent) random $r_t$ values is generated, with its size equal to the maximum time-steps considered for the CQW dynamics (say, $t=15$). It represents one disorder realization, and subsequently, the realization average is computed with 500 such realizations for estimating a quantity $\Omega_4(t)$ like, $E_{av}$ and $P_{av}(x=0)$, i.e.,  

\begin{align}
\begin{split} 
\langle \Omega_4(t) \rangle=\frac{1}{500}\sum_{j=1}^{500}\Omega_4 (\ket{\psi(t)}_j\bra{\psi(t)}_j)\;.\;
\end{split}
\label{eq13}
\end{align}


\begin{figure}[h!]
\includegraphics[width=18cm,height=8.5cm]{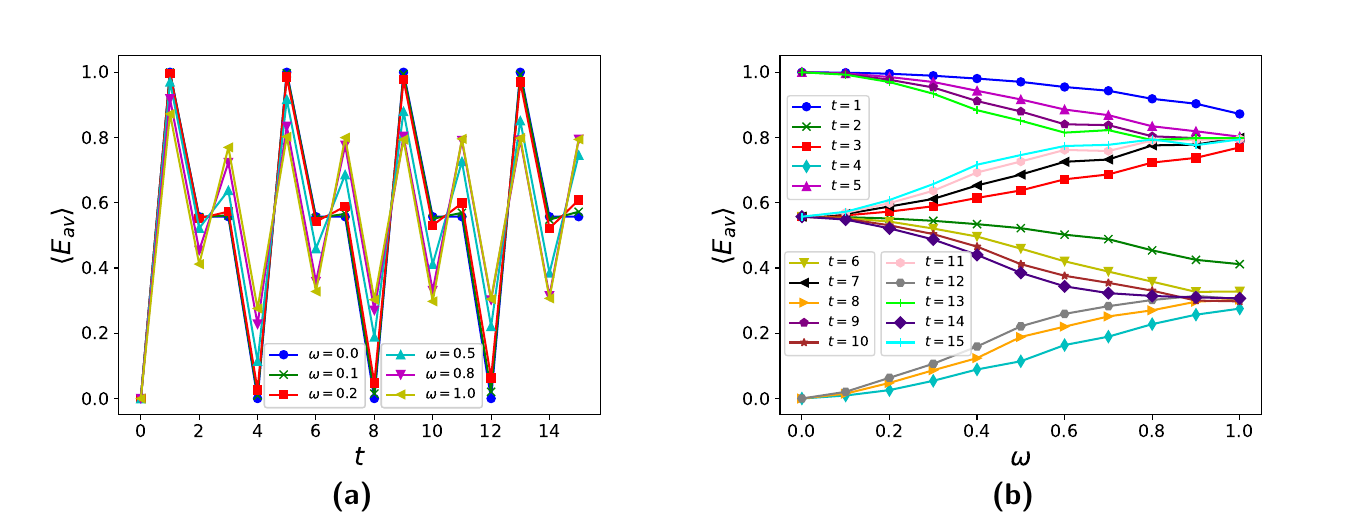}
\caption{(a) SPE $\langle E_{av} \rangle$ vs. time step $t$ for different dynamic-coin-disorder strength ($\omega$); (b) SPE $\langle E_{av} \rangle$ vs. dynamic-coin-disorder strength $\omega$ for different time-steps $t$ for 4-cycle.}
\label{f9}
\end{figure}

\begin{figure}[h!]
\includegraphics[width=18cm,height=8.5cm]{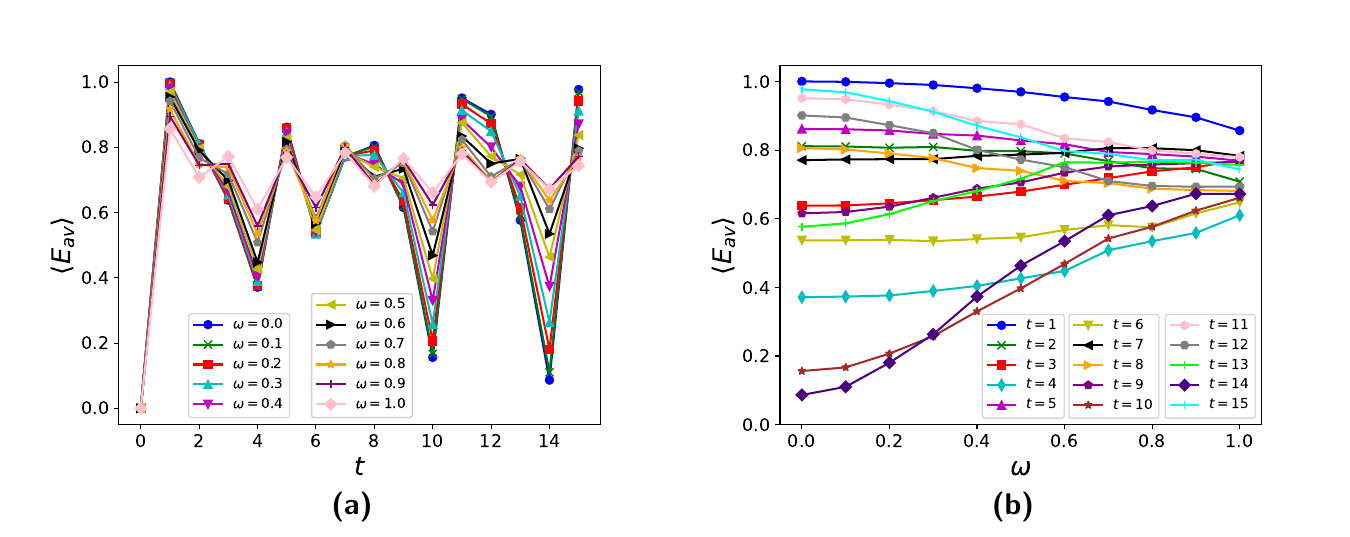}
\caption{(a) SPE $\langle E_{av} \rangle$ vs. time step $t$ for different dynamic-coin-disorder strength ($\omega$); (b) SPE $\langle E_{av} \rangle$ vs. dynamic-coin-disorder  strength $\omega$ for different time-steps $t$ for 3-cycle.}
\label{f10}
\end{figure}

The quNit-qubit SPE dynamics show significant differences under dynamic coin disorder for different disorder strengths, i.e., for $\omega$ ranging from 0 (clean-CQW) to 1, see Figs.~\ref{f9} and \ref{f10}. 
MESPS at $t=1$ remains unperturbed for both 3- and 4-cycles and for any arbitrary disorder strengths, for two-phase symmetric initial states, i.e., main text Eq.~(1) with $\theta=\frac{\pi}{2}$ with $\phi=\frac{\pi}{2},\frac{3\pi}{2}$, i.e., $\frac{\ket{0_p,0_c}\pm i\ket{0_p,1_c}}{\sqrt{2}}$, see below Subsec.~4 for its derivation. From Fig.~\ref{f9}(b), one can observe that SPE remains almost unaffected by small disorder strengths $\omega$; see, for example, SPE values at $t=1,3,5,7,9,...$ and the SPE increases under the coin disorder too, see for example SPE values at $t=3,4,7,8,11,12,15,...$. Moreover, we observe that the dynamic coin disorder revives single-particle entanglement from zero, i.e., from an otherwise separable state observed at $t=4,8,12$, see  Fig.~\ref{f9}(b) (cyan, yellow, grey). We find similar results for the 3-cycle case, see Fig.~\ref{f10}, i.e., SPE is almost unaffected by dynamic coin disorders of small strength. Further, we observe that the dynamic coin disorder can increase SPE at specific time steps for 3-cycle, e.g., see SPE at $t=4,6,10,13,14$, in  Fig.~\ref{f10}.

\subsubsection{Static vs dynamic coin disorder}
Static and dynamic coin disorder have many similar effects on CQW and generated SPE values, such as both preserving odd-even parity (4-cycle), obeying physical time-cone (all cycles), increasing SPE (at specific time steps $t$) as well as reviving SPE from zero. However, static coin disorder, on average, is the more substantial coin disorder; see SPE values at $t=3,5,7,11,13,15$ in Figs.~\ref{f7} and \ref{f9} for 4-cycle, where we see the static coin disorder reduces SPE more than the dynamic variant. A similar result can also be observed for the 3-cycle; see Figs.~\ref{f8} and \ref{f10}.

\subsubsection{Robustness of MESPS at $t=1$ for coin disorder in both 3 and 4-cycles}
\label{appb}
Herein, we provide detailed proof below that MESPS at time-step $t=1$ is unperturbed against static coin disorder of arbitrary strength for phase symmetric initial states, i.e., 
\begin{equation}
\ket{\psi(0)} =\frac{\ket{0_p,0_c}+ i\ket{0_p,1_c}}{\sqrt{2}},\;\frac{\ket{0_p,0_c}- i\ket{0_p,1_c}}{\sqrt{2}},
\label{eq30}
\end{equation}
which are quantum states in Eq.~(\ref{eq21}) for $\theta=\frac{\pi}{2}$ with $\phi=\frac{\pi}{2}$ and $\phi=\frac{3\pi}{2}$.
To analytically show this robustness for static coin disorder, we start with the time-evolved state at $t=1$,
\begin{equation}
    \ket{\psi(t=1)}=\hat{S}(t=1)\cdot[\sum_{x}\ket{x}\bra{x}\otimes \hat{C}(\rho(x))]\ket{\psi(0)}\;,
\end{equation} 
where the translation operator is, 
\begin{align}
\begin{split} \hat{S}(t=1) =& \sum_{s=0}^{1}\sum_{x=0}^{k-1}\ket{((x+2s-1) \text{ mod } k)_p}\bra{x_p}\otimes\ket{s_c}\bra{s_c},
\end{split}
\label{eq32}
\end{align} and the coin operator at initial walker site $\ket{x_p=0_p}$ is $\hat{C}(\rho(x=0))=\begin{pmatrix}
\sqrt{\rho(x=0)} & \sqrt{1-\rho(x=0)}\\
\sqrt{1-\rho(x=0)} & -\sqrt{\rho(x=0)}
\end{pmatrix}$, with $\rho(x=0)=\frac{1}{2}(1+\omega r_0)=\rho_0 \text{ (say)}$, and  $r_0\in[1,-1]$ are random numbers drawn from a uniform distribution~\cite{sasha}. The evolved state now becomes, 
 \begin{equation}
\ket{\psi(t=1)}=(\sqrt{\rho_0}\cos(\frac{\theta}{2})+\sqrt{1-\rho_0}e^{i\phi}\sin(\frac{\theta}{2}))\ket{(k-1)_p,0_c}+(\sqrt{1-\rho_0}\cos(\frac{\theta}{2})-\sqrt{\rho_0}e^{i\phi}\sin(\frac{\theta}{2})) \ket{1_p,0_c}. 
\label{eq33}
\end{equation}

\begin{figure}[h!]
\includegraphics[width=18cm,height=7cm]{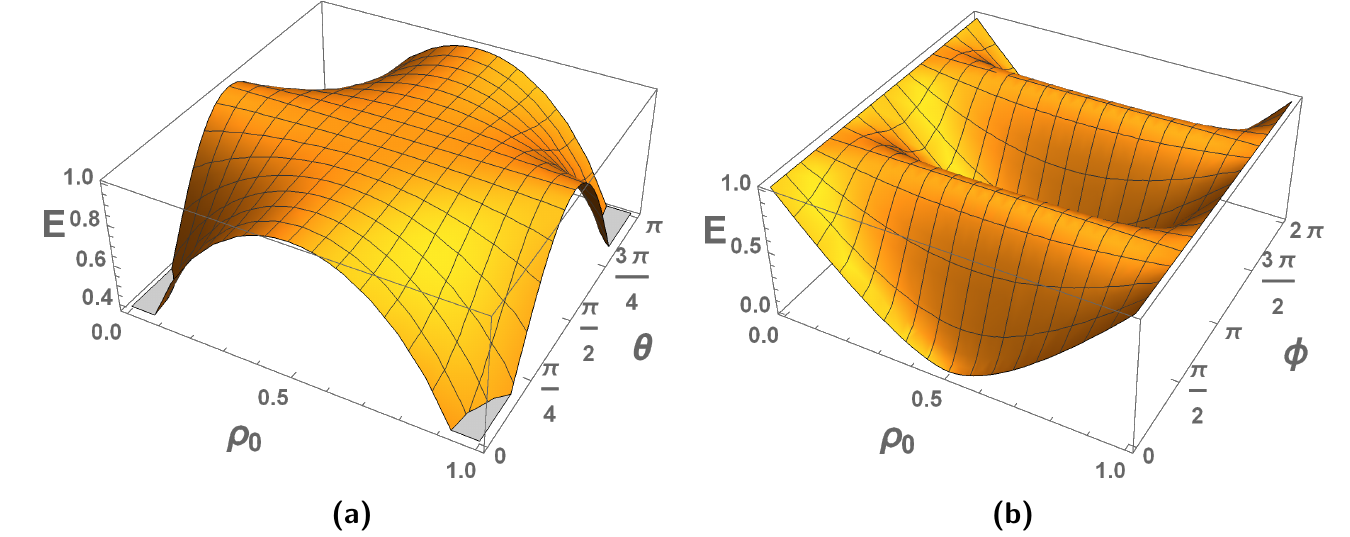}
\caption{(a) SPE $ E $ at time step $t=1$ as a function of randomised coin-parameter $\rho_0$ and initial state parameter $\theta$ for $\phi=\frac{\pi}{2}$; (b) SPE $ E $ at time step $t=1$ as a function of randomised coin-parameter $\rho_0$ and initial state phase parameter $\phi$ for initial state parameter $\theta=\frac{\pi}{2}$. This concludes MESPS ($ E =1$ ) is unaffected for any random values of $\rho_0$ for $\phi=\frac{\pi}{2},\frac{3\pi}{2}$ with $\theta=\frac{\pi}{2}$. }
\label{f13}
\end{figure}

This state for 
$\phi=\frac{\pi}{2},$ or $\frac{3\pi}{2}$, yields the reduced density matrix, 

\begin{equation}
\rho_c(t)=\text{Tr}_p(\ket{\psi(1)}\bra{\psi(1)})=\begin{pmatrix}
\sin^2(\frac{\theta}{2})+\sqrt{\rho_0}\cos(\theta) & 0\\
0 & \cos^2(\frac{\theta}{2})-\sqrt{\rho_0}\cos(\theta)
\end{pmatrix},
\label{eq34}
\end{equation}with the diagonal elements its eigenvalues, and therefore
at $\theta=\frac{\pi}{2}$, it results in a maximally mixed reduced density matrix, $\rho_c(t)=\frac{1}{2}\begin{pmatrix}
1 & 0\\
0 & 1
\end{pmatrix},$ and in turn yields entanglement entropy, 
$E(\rho_c)=1$ for any random $\rho_0$ values, i.e., for the static coin disorder of arbitrary strength $\omega$. It is also shown in Fig.~\ref{f13}(a-b).

Thus, static coin disorder does not affect MESPS at $t=1$ for the two phase-symmetric initial states given in Eq.~(\ref{eq30}). This exciting result holds for any disorder strength $\omega$, as the reduced density matrix $\rho_c(t)$ is entirely independent of $\rho_0$ and hence $\omega$.

A similar proof of the robustness of MESPS (at $t=1$) against dynamic coin disorder of arbitrary strength can also be derived for the same phase symmetric initial state shown in Eq.~(\ref{eq30}), where the time-evolved state at $t=1$ is,
\begin{equation}
    \ket{\psi(t=1)}=\hat{S}(t=1)\cdot[\mathbf{I}_p\otimes \hat{C}(\rho(t=1))]\ket{\psi(0)}\;,
\label{dyncoin}
\end{equation} 
with the translation (shift) operator, 
\begin{align}
\begin{split} \hat{S}(t=1) =& \sum_{s=0}^{1}\sum_{x=0}^{k-1}\ket{((x+2s-1) \text{ mod } k)_p}\bra{x_p}\otimes\ket{s_c}\bra{s_c},
\end{split}
\label{eq32}
\end{align} and the coin operator, $\hat{C}(\rho(t=1))=\begin{pmatrix}
\sqrt{\rho(t=1)} & \sqrt{1-\rho(t=1)}\\
\sqrt{1-\rho(t=1)} & -\sqrt{\rho(t=1)}
\end{pmatrix}$, with $\rho(t=1)=\frac{1}{2}(1+\omega r_1)=\rho'_0 \text{ (say)}$, and $r_1\in[1,-1]$ are random numbers drawn from a uniform distribution~\cite{sasha}. The evolved state at $t=1$ now becomes, 
 \begin{equation}
\ket{\psi(t=1)}=(\sqrt{\rho'_0}\cos(\frac{\theta}{2})+\sqrt{1-\rho'_0}e^{i\phi}\sin(\frac{\theta}{2}))\ket{(k-1)_p,0_c}+(\sqrt{1-\rho'_0}\cos(\frac{\theta}{2})-\sqrt{\rho'_0}e^{i\phi}\sin(\frac{\theta}{2})) \ket{1_p,0_c}. 
\label{eq33}
\end{equation} 

This state for 
$\phi=\frac{\pi}{2},$ or $\frac{3\pi}{2}$, and $\theta=\frac{\pi}{2}$, yields the reduced density matrix (similar to the static-coin disorder case), $\rho_c(t)=\frac{1}{2}\begin{pmatrix}
1 & 0\\
0 & 1
\end{pmatrix},$ which is  maximally mixed. Thus this, too yields entanglement entropy, 
$E(\rho_c)=1$ for any random $\rho'_0$ values, i.e., for the dynamic coin disorder of arbitrary strength $\omega$. Therefore, this proves that dynamic coin disorder does not affect MESPS at $t=1$ for the two phase-symmetric initial states given in Eq.~(\ref{eq30}).


\subsection{Position disorder and SPE for 3-cycle}
\label{appc}
\begin{figure}[h!]
\includegraphics[width=18cm,height=8.5cm]{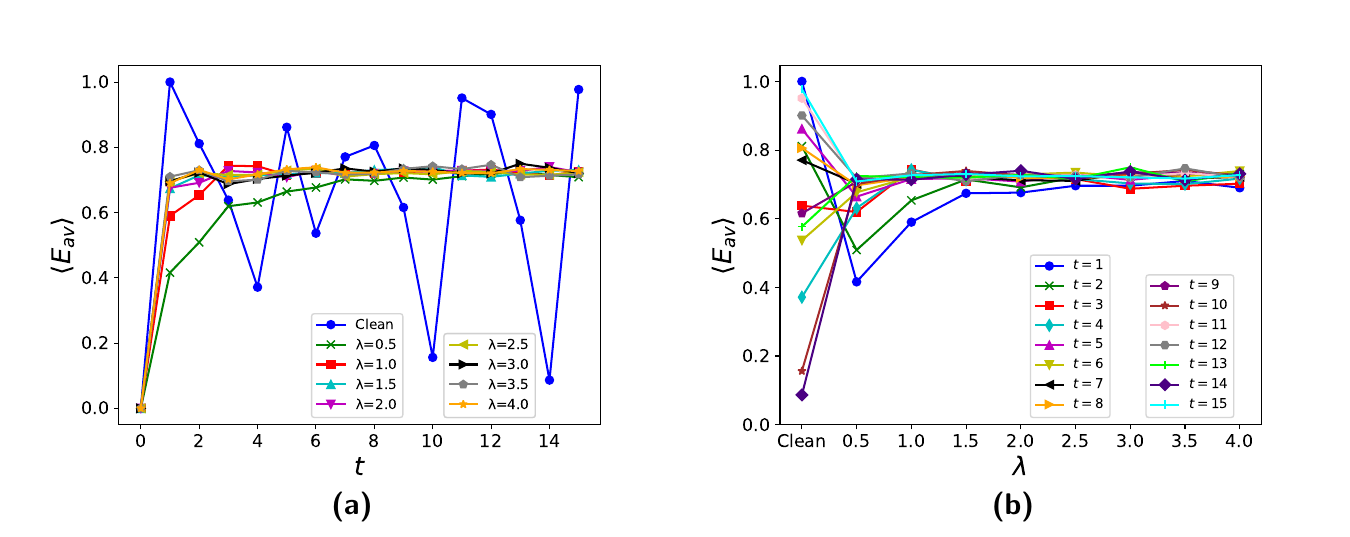}
\caption{(a) SPE $\langle E_{av} \rangle$ vs. time step $t$ for different position-disorder strength ($\lambda$); (b) SPE $\langle E_{av} \rangle$ vs. position-disorder strength $\lambda$ for different time-steps $t$ for 3-cycle.}
\label{f12}
\end{figure}

As discussed in the main text, the quNit-qubit SPE is highly vulnerable to position disorder compared to phase or coin disorder. The recurrence or periodicity of the SPE (or MESPS for clean-CQW) in 4-cycle CQW is distorted under position disorder. Moreover, under position disorder, SPE saturates to a fixed value in 3-cycle as was also seen in the 4-cycle case (see main text Fig.~(4)), irrespective of the disorder strength $\lambda$ at large $t$; see Fig.~\ref{f12}. 
Interestingly, in line with other disorders, a position disorder can also increase SPE at some time steps; see SPE at time-steps $t=4,6,9,10,13,14$ in Fig.~\ref{f12} for 3-cycle.

\newpage
\subsection{Algorithm to see effects of disorders via CQW on SPE}
Herein, we put forth our algorithm to implement the dynamic phase disorder in a CQW to study its effect on SPE, as follows:
\begin{algorithmic}[1]
\Require time-step ($N$), $simulations$ (500), $\ket{\psi(0)}$, disorder\_strength
\Return average\_SPE  (i.e., $\langle E_{av} \rangle$ )
\State Initialize  average\_SPE as an array of zeros of length $N$
\For{$simu = 1$ to $simulations$}
    \State \textbf{Generate} random\_$\alpha_{0}$ using uniform distribution with parameter disorder\_strength
    \State Initialize $\ket{\psi(N)}$ as $\ket{\psi(0)}$
    \State Initialize avg\_SPE as an array of zeros of length $N$
    \For{$step = 1$ to $N$}
       \For{{$run= 1$ to $step$}}
        \State $\alpha_{0} \gets$ random\_$\alpha_{0}$[$run$]
        \State Compute translation matrix  $\hat{S}_{dp}(run)$     
        \State Compute $U_{k}(run)$ using $\hat{S}_{dp}(run)$ and $\hat{H}$ coin
        \State Update $\ket{\psi(N)}$ using $U_{k}(run)$
        \State Calculate the reduced density matrix $\rho_c$ from $\ket{\psi(N)}$
        \State Calculate eigenvalues of $\rho_c$
        \State Calculate SPE from the eigenvalues
        \State Integrate SPE over $\theta$ from $0$ to $\pi$ to obtain the average SPE$_{av}$ (i.e., $E_{av}$)
        
    \EndFor
    \State Update avg\_SPE[$step$] with the computed SPE
    \EndFor
    \State Update average\_SPE=avg\_SPE+average\_SPE
\EndFor
\State Calculate average\_SPE by dividing by $simulations$
\For{$step = 1$ to $N$}
    \State Print the average SPE value average\_SPE[$step$] at $step$
\EndFor
\end{algorithmic}

Similarly, algorithms for implementing static phase disorder, coin disorder (static and dynamic), and position disorder can also be written, following the main text. These algorithms can be implemented in Python.
\newpage
\subsection{Comparison between different approaches to generate SPE with disorder and study disorder effects on SPE}
\label{appd}
In Table~\ref{tab1}, we have compared our results with some prior works; see Refs.~\cite{sasha,prl,china, ntuyong,naves22} on generating SPE in DTQW with 1D line. Refs.~\cite{sasha,prl,china,ntuyong,naves22} focus on generating SPE using disorder (each work considered either a coin, phase, or position disorder) but only for some specific initial states. These proposals do not look at the effects of a disorder on MESPS or SPE (via clean QW). We bridge this gap and comprehensively report on generating SPE via disorder. We also discuss the effects of various disorders on SPE's enhancement, revival, and resilience under phase, coin, and position disorder of arbitrary strength with the most general initial state.
\begin{table*}[h!]
\centering
\newpage
\caption{SPE vs. disorder  (comparison between this work and other approaches)}
\label{tab7}
\resizebox{\textwidth}{!}{%
\begin{tabular}{|c|l|l|l|l|l|l|l|}
\hline
\textbf{Properties$\downarrow$/Model$\rightarrow$} &
\multicolumn{1}{c|}{\textbf{\begin{tabular}[c]{@{}c@{}} Probing resilience of SPE \\under coin, position and phase\\disorders(This paper)\\ \textcolor{blue}{(DTQW on cyclic graphs)} \end{tabular}}} &
\multicolumn{1}{c|}{\textbf{\begin{tabular}[c]{@{}c@{}}SPE via static\\ coin disorder~\cite{sasha}\\ \textcolor{blue}{(DTQW on 1D line)} \end{tabular}}} &
\multicolumn{1}{c|}{\textbf{\begin{tabular}[c]{@{}c@{}}SPE via \\phase disorder~\cite{ntuyong}\\ \textcolor{blue}{(DTQW on 1D line)} \end{tabular}}} &
\multicolumn{1}{c|}{\textbf{\begin{tabular}[c]{@{}c@{}}SPE via dynamic\\ coin disorder~\cite{prl}\\ \textcolor{blue}{(DTQW on 1D line)} \end{tabular}}} 
&
\multicolumn{1}{c|}{\textbf{\begin{tabular}[c]{@{}c@{}} SPE via aperiodic\\ coin disorder~\cite{china} \\ \textcolor{blue}{(DTQW on 1D line)}  \\ \end{tabular}}} & \multicolumn{1}{c|}{\textbf{\begin{tabular}[c]{@{}c@{}} SPE via noisy\\ position disorder~\cite{naves22} \\ \textcolor{blue}{(DTQW on 1D line)}  \\ \end{tabular}}} \\ \hline
\hline
\textbf{\begin{tabular}[c]{@{}c@{}}Initial state(s)
\end{tabular}} &
\begin{tabular}[c]{@{}l@{}}Most general initial\\ state, see main text Eq.~(1). \\All possible (infinite number\\ of) initial states are considered.
\end{tabular} &
\begin{tabular}[c]{@{}l@{}}A single specific\\ initial state is considered.\end{tabular} &
\begin{tabular}[c]{@{}l@{}}Only 2016 initial states \\are considered when \\averaging for SPE value.\end{tabular} &
\begin{tabular}[c]{@{}l@{}}A single specific\\ initial state is considered.\end{tabular} &
\begin{tabular}[c]{@{}l@{}}A single specific\\ initial state is considered.\end{tabular} &
\begin{tabular}[c]{@{}l@{}}Specific initial state\\ and states with\\ specific phase ($\phi$) \\are considered.\end{tabular} \\ \hline
\textbf{\begin{tabular}[c]{@{}c@{}}Disorders studied
\end{tabular}} &
\begin{tabular}[c]{@{}l@{}}All kinds: coin (static and dynamic)\\phase 
(static and dynamic) \\and position disorder.
\end{tabular} &
\begin{tabular}[c]{@{}l@{}}Only static \\coin disorder.\end{tabular} &
\begin{tabular}[c]{@{}l@{}}Only phase\\ disorder.\end{tabular} &
\begin{tabular}[c]{@{}l@{}}Only dynamic\\ coin disorder.\end{tabular}  &
\begin{tabular}[c]{@{}l@{}}Only aperiodic coin\\ disorder from sequences:\\ Fibonacci, Thue-Morse.\end{tabular} &
\begin{tabular}[c]{@{}l@{}}Only noisy time-\\dependent shift disorder\\ from $q$-exponential \\distribution.\end{tabular} \\ \hline
\textbf{\begin{tabular}[c]{@{}c@{}}Disorder category
\end{tabular} }&
\begin{tabular}[c]{@{}l@{}}Nondeterministic: Poissonian and \\uniform distributions of\\ different disorder strengths. \end{tabular} &
\begin{tabular}[c]{@{}l@{}}Nondeterministic: \\uniform distribution. \end{tabular} &
\begin{tabular}[c]{@{}l@{}}Nondeterministic: Binary \\and uniform distributions. \end{tabular} &
\begin{tabular}[c]{@{}l@{}}Nondeterministic: \\continuous uniform\\ distribution.\\ \end{tabular} &
\begin{tabular}[c]{@{}l@{}}Deterministic: coins \\from simple sequences.\end{tabular} &
\begin{tabular}[c]{@{}l@{}}Non-deterministic: shift\\ parameter drawn from\\ q-exponential distribution.\end{tabular} \\  \hline
\textbf{\begin{tabular}[c]{@{}c@{}}\textit{Odd-even} parity\\and physical \textit{time-cone}.
\end{tabular}} &
\begin{tabular}[c]{@{}l@{}}Preserved under coin and phase\\ disorder (both static and dynamic)\\ and broken in position\\ disorder, leading to larger \\spreading (useful in quantum\\ algorithms and quantum simulators).
\end{tabular} &
\begin{tabular}[c]{@{}l@{}}Preserved under\\ coin disorder.\end{tabular} &
\begin{tabular}[c]{@{}l@{}}Not discussed.\end{tabular} &
\begin{tabular}[c]{@{}l@{}}Not discussed.\end{tabular} &
\begin{tabular}[c]{@{}l@{}}Not discussed.\end{tabular} &
\begin{tabular}[c]{@{}l@{}}Not discussed.\end{tabular} \\ 
\hline
\textbf{\begin{tabular}[c]{@{}c@{}}Resilience of SPE \\(or MESPS) under \\coin disorder\end{tabular}} &
\begin{tabular}[c]{@{}l@{}}Yes.
Coin disorder case:\\ SPE (and maximal SPE)\\ reduces insignificantly for\\ small disorder strength.\\ Thus, SPE is resilient to\\ small coin-disorder strength.\\ Further increasing\\ disorder strength, SPE \\decreases slowly. Interestingly, for\\moderate to large disorder\\ strengths, SPE again increases\\ at certain time steps.\end{tabular} 
&
\begin{tabular}[c]{@{}l@{}}Not discussed.\end{tabular} &
\begin{tabular}[c]{@{}l@{}}Not discussed.\end{tabular} &
Not discussed. &

\begin{tabular}[c]{@{}l@{}}Not discussed.\end{tabular} &
\begin{tabular}[c]{@{}l@{}}Not discussed.\end{tabular} \\ 
\hline
\textbf{\begin{tabular}[c]{@{}c@{}}Resilience of SPE \\(or MESPS) under \\phase disorder\\ \end{tabular}} &
\begin{tabular}[c]{@{}l@{}}Yes.
Phase disorder case: \\SPE increases for small to \\ significant disorder strengths at \\ specific time steps.
\end{tabular} 
&
\begin{tabular}[c]{@{}l@{}}Not discussed.\end{tabular} &
\begin{tabular}[c]{@{}l@{}}Not discussed.\end{tabular} &
Not discussed. &
\begin{tabular}[c]{@{}l@{}}Not discussed.\end{tabular} &
\begin{tabular}[c]{@{}l@{}}Not discussed.\end{tabular} \\ 
\hline
\textbf{\begin{tabular}[c]{@{}c@{}}Resilience of SPE \\(or MESPS) under \\position disorder\\ \end{tabular}} &
\begin{tabular}[c]{@{}l@{}}No. However, under position\\ disorder of arbitrary \\ strength$(\lambda)$ SPE saturates\\ at all time steps,\\ unlike coin and phase disorders.
\end{tabular} 
&
\begin{tabular}[c]{@{}l@{}}Not discussed.\end{tabular} &
\begin{tabular}[c]{@{}l@{}}Not discussed.\end{tabular} &
Not discussed. &
\begin{tabular}[c]{@{}l@{}}Not discussed.\end{tabular} &
\begin{tabular}[c]{@{}l@{}}Not discussed.\end{tabular} \\ 
\hline

\textbf{\begin{tabular}[c]{@{}c@{}} Revival of SPE under \\all kinds of disorder\\ \end{tabular}} &
\begin{tabular}[c]{@{}l@{}}Yes. With small to large \\ strengths of coin or phase \\or position disorder, SPE \\is generated at time steps where\\ there was no SPE in \\clean-QW for \\the 4-cycle.\end{tabular} 
&
\begin{tabular}[c]{@{}l@{}}Not discussed.\end{tabular} &
\begin{tabular}[c]{@{}l@{}}Not discussed.\end{tabular} &
Not discussed. &
\begin{tabular}[c]{@{}l@{}}Not discussed.\end{tabular} &
\begin{tabular}[c]{@{}l@{}}Not discussed.\end{tabular} \\ 
\hline
\textbf{\begin{tabular}[c]{@{}c@{}}Robustness of MESPS \\(maximal SPE) at \\time-step 1 against\\ different disorders \end{tabular}} &
\begin{tabular}[c]{@{}l@{}}Yes. At $t=1$, MESPS (maximal\\ SPE value of 1)\\ is robust against dynamic and\\ static phase disorders for\\ arbitrary initial states. At the same time, \\it is guaranteed for two specific \\phase-symmetric initial states in \\the case of static and dynamic\\ coin disorder. These MESPS (maximal \\SPE) preserved against\\ disorders are analytically proved \\(in main text and \\supplementary material). \end{tabular} 
&
\begin{tabular}[c]{@{}l@{}}Not achieved.\end{tabular} &
\begin{tabular}[c]{@{}l@{}}Not achieved.\end{tabular} &
Not achieved. &
\begin{tabular}[c]{@{}l@{}}Not discussed.\end{tabular} &
\begin{tabular}[c]{@{}l@{}}Not discussed.\end{tabular} \\ 
\hline
\textbf{\begin{tabular}[c]{@{}c@{}}Controlling SPE by \\tuning the disorder of \\certain strengths  \end{tabular}} &
\begin{tabular}[c]{@{}l@{}}Yes, coin/phase/position disorder\\ of appropriate strengths can reduce\\ as well as increase SPE.\\ We focused on both generation \\of SPE from separable states\\and control of the MESPS and SPE \\values via disorder strengths.
\end{tabular} 
&
\begin{tabular}[c]{@{}l@{}}Not discussed.\end{tabular} &
\begin{tabular}[c]{@{}l@{}}Not discussed. It focuses\\ only on the \\generation of high SPE\\ from product states.\end{tabular} &
\begin{tabular}[c]{@{}l@{}}Not discussed. It focuses\\ only on the \\generation of high SPE\\ from product states.\end{tabular}&
\begin{tabular}[c]{@{}l@{}}Not discussed. It focuses\\ only on the \\generation of high SPE\\ from product states.\end{tabular} &
\begin{tabular}[c]{@{}l@{}}Not discussed. It focuses\\ only on the \\generation of high SPE\\ from product states.\end{tabular} \\
\hline

\end{tabular}%
}
\label{tab1}
\end{table*}
\newpage
\twocolumngrid

\newpage
\twocolumngrid


\begin{thebibliography}{1}

\bibitem{wineland}C. Monroe, D. M. Meekhof, B. E. King, D. J. Wineland, A “Schrödinger Cat” Superposition State of an Atom, Science, 272, 1131 (1996).
\bibitem{aqs} S. Azzini, S. Mazzucchi, V. Moretti, D. Pastorello, and L.
Pavesi, Single-Particle Entanglement, Advanced Quantum Technologies 3, 2000014 (2020).
\bibitem{fang}	Xiao-Xu Fang, Kui An, Bai-Tao Zhang, Barry C. Sanders, He Lu, Maximal coin-position entanglement generation in a quantum walk for the third step and beyond regardless of the initial state, Phys. Rev. A 107, 012433 (2023).
 
\bibitem{p2}Dinesh Kumar Panda and Colin Benjamin, Recurrent generation of maximally entangled single-particle states via quantum walks on cyclic graphs, Phys. Rev. A 108, L020401 (2023).
\bibitem{gratsea_lewenstein_dauphin_2020} A. Gratsea, M. Lewenstein, and A. Dauphin, Generation of hybrid maximally entangled states in a one-dimensional quantum walk, Quantum Science and Technology {5}, 025002 (2020).
\bibitem{qjoining}  C. Vitelli, N. Spagnolo, L. Aparo, F. Sciarrino, E. Santamato, and L. Marrucci, Joining the quantum state of two photons into one, Nature Photonics 7, 521 (2013).
\bibitem{p4} Dinesh Kumar Panda and Colin Benjamin, Quantum cryptographic protocols with dual messaging system via 2D alternate quantum walk of a genuine single-photon entangled state, J. Phys. A: Math. Theor. 58 01LT01 (2025).
\bibitem{gmeuse} R. Raussendorf and H. J. Briegel, Phys. Rev. Lett. 86, 5188 (2001).
\bibitem{gmeuse20}M. Blasone, F. Dell’Anno, S. De Siena, M. Di Mauro, and F. Illuminati, Multipartite entangled states in particle mixing, Phys. Rev. D 77, 096002 (2008).
\bibitem{gmeuse2}Abhishek Kumar Jha, Supratik Mukherjee, and Bindu A. Bambah, Tri-partite entanglement in neutrino oscillations, Mod. Phys. Lett. A 36, 2150056 (2021).
\bibitem{qudit1}Kumel H. Kagalwala, Giovanni Di Giuseppe, Ayman F. Abouraddy and Bahaa E. A. Saleh, Single-photon three-qubit quantum logic using spatial light modulators, Nature Communications 8, 739 (2017).
\bibitem{qudit2}Poolad Imany, Navin B. Lingaraju, Mohammed S. Alshaykh,  Daniel E. Leaird, and Andrew M. Weiner, Probing quantum walks through coherent control of high-dimensionally entangled photons, Science Advances 6, eaba8066 (2020).

\bibitem{p1}Dinesh Kumar Panda, B. Varun Govind and Colin Benjamin, Generating highly entangled states via discrete-time quantum
walks with Parrondo sequences, Physica A 608, 128256 (2022).
\bibitem{qw93}Y. Aharonov, L. Davidovich, and N. Zagury, Quantum random
walks, Phys. Rev. A 48, 1687 (1993).
\bibitem{p3}Dinesh Kumar Panda and Colin Benjamin, Designing three-way-entangled and nonlocal two-way-entangled single-particle states via alternate quantum walks, Phys. Rev. A 111, 012420 (2025).
\bibitem{cold1}
Perets, H. B. et al. Realization of quantum walks with negligible decoherence
in waveguide lattices. Phys. Rev. Lett. 100, 170506 (2008).
\bibitem{karski}Michal Karski et al., Quantum Walk in Position Space with Single Optically Trapped Atoms, Science 325, 174-177(2009).
\bibitem{bian} Z.-H. Bian, J. Li, X. Zhan, J. Twamley, and P. Xue, Experimental implementation of a quantum walk on a circle with single
photons, Phys. Rev. A 95, 052338 (2017).

\bibitem{ph1}Peruzzo, A. et al. Quantum walks of correlated photons. Science 329,
1500–1503 (2010).
\bibitem{ph2}Schreiber, A. et al. Photons walking the line: a quantum walk with adjustable
coin operations. Phys. Rev. Lett. 104, 050502 (2010).
\bibitem{ph3} Broome, M. A. et al. Discrete single-photon quantum walks with tunable
decoherence. Phys. Rev. Lett. 104, 153602 (2010).
\bibitem{ph4} Tamura, M., Mukaiyama, T. \& Toyoda, K. Quantum walks of a phonon in
trapped ions. Phys. Rev. Lett. 124, 200501 (2020).
\bibitem{ion1}Schmitz, H. et al. Quantum walk of a trapped ion in phase space. Phys. Rev.
Lett. 103, 090504 (2009).
\bibitem{ion2} Zähringer, F. et al. Realization of a quantum walk with one and two trapped
ions. Phys. Rev. Lett. 104, 100503 (2010).
\bibitem{nmrqw} Ryan, C. A., Laforest, M., Boileau, J. C. \& Laflamme, R. Experimental
implementation of a discrete-time quantum random walk on an NMR
quantum-information processor. Phys. Rev. A 72, 062317 (2005).

\bibitem{sasha}Louie Hong Yao and Sascha Wald, Coined quantum walks on the line: Disorder, entanglement, and localization, Phys. Rev. E 108, 024139 (2023).

\bibitem{prl}Rafael Vieira, Edgard P. M. Amorim, and Gustavo Rigolin, Dynamically Disordered Quantum Walk as a Maximal Entanglement Generator, Phys. Rev. Lett. 111, 180503 (2013).

\bibitem{noise3} J. Marino and S. Diehl, Driven Markovian quantum criticality,
Phys. Rev. Lett. 116, 070407 (2016).
\bibitem{noise2}K. Seetharam, A. Lerose, R. Fazio, and J. Marino, Correlation engineering via nonlocal dissipation, Phys. Rev. Res. 4,
013089 (2022).
\bibitem{noise4} F. Verstraete, M. M. Wolf, and J. Ignacio Cirac, Quantum computation and quantum-state engineering driven by dissipation,
Nat. Phys. 5, 633 (2009).
\bibitem{noise1}
Cristiano Muzzi, Mikheil Tsitsishvili and Giuliano Chiriacò, Entanglement enhancement induced by noise in inhomogeneously monitored systems, Phys. Rev. B 111, 014312 (2025).
\bibitem{noise7}Z. Cai, R. Babbush, S. C. Benjamin, S. Endo, W. J. Huggins,
Y. Li, J. R. McClean, and T. E. O’Brien, Quantum error mitigation, Rev. Mod. Phys. 95, 045005 (2023).

\bibitem{noise6} A. Lowe, M. H. Gordon, P. Czarnik, A. Arrasmith, P. J. Coles,
and L. Cincio, Unified approach to data-driven quantum error
mitigation, Phys. Rev. Res. 3, 033098 (2021).
\bibitem{noise5}  A. Kandala, K. Temme, A. D. Córcoles, A. Mezzacapo, J. M.
Chow, and J. M. Gambetta, Error mitigation extends the computational reach of a noisy quantum processor, Nature (Lond.)
567, 491 (2019).
\bibitem{supple} See Supplemental Material at [Link to be provided by Publisher] for more details, which includes Refs.~\cite{p2,p4,prl,sasha,naves22,eqw1,eqw2,portugal, ntuyong, china}.
\bibitem{viera}R. Vieira, E. P. M. Amorim, and G. Rigolin, Dynamically Disordered Quantum Walk as a Maximal Entanglement Generator,
Phys. Rev. Lett. 111, 180503 (2013).
\bibitem{janzing_2009}D. Janzing, Entropy of entanglement, Compendium of Quantum Physics, 205–209 (2009).
\bibitem{ntuyong}Meng Zeng and Ee Hou Yong, Discrete-Time Quantum Walk with Phase Disorder: Localization and Entanglement Entropy. Sci Rep 7, 12024 (2017).
\bibitem{aditi19} 
Sreetama Das, Shiladitya Mal, Aditi Sen(De), and Ujjwal Sen, Inhibition of spreading in quantum random walks due to quenched Poisson-distributed disorder, Phys. Rev. A 99, 042329 (2019).
\bibitem{pqc}Daniel J. Bernstein and Tanja Lange, Post-quantum cryptography,
Nature volume 549, pages188–194 (2017).
\bibitem{naves22}Caio B. Naves, Marcelo A. Pires, Diogo O. Soares-Pinto, and Sílvio M. Duarte Queirós, Enhancing entanglement with the generalized elephant quantum walk from localized and delocalized states, Phys. Rev. A 106, 042408 (2022).

\bibitem{eqw1}Marcelo A. Pires, Giuseppe Di Molfetta and Sílvio M. Duarte Queirós, Multiple transitions between normal and hyperballistic diffusion in quantum walks with time-dependent jumps, Sci Rep 9, 19292 (2019).
\bibitem{eqw2} Di Molfetta, G., Soares-Pinto, D. O. \& Queirós, S. M. D., Elephant quantum walk. Phys. Rev. A 97, 062112 (2018).
\bibitem{portugal}R. Portugal, Quantum Walks and Search Algorithms (Springer,
New York, 2013).
\bibitem{china}Ting-Ting Liu et al., The entanglement of deterministic aperiodic
quantum walks, Chinese Phys. B 27 120305 (2018).
\bibitem{Chandra2022} P. A. Ameen Yasir and C. M. Chandrashekar, Generation of hyperentangled states and two-dimensional quantum walks using J or  q plates and polarization beam splitters, Phys. Rev. A 105, 012417 (2022).
\bibitem{expt1-science} M. Karski et al., Quantum Walk in Position Space with Single Optically Trapped Atoms, Science 325, 174 (2009).
\bibitem{jplate}Robert C. Devlin, Antonio Ambrosio, Noah A. Rubin, J. P. Balthasar Mueller, Federico Capasso, Arbitrary spin-to–orbital angular momentum conversion of light, Science 358, 896-901 (2017).
\bibitem{2dqw-expt}Hao Tang et al. ,Experimental two-dimensional quantum walk on a photonic chip, SCIENCE ADVANCES 4, eaat3174 (2018).
\end{thebibliography}
\end{document}